\documentclass[%
 reprint,
 superscriptaddress,
 amsmath,amssymb,
 aps,
]{revtex4-2}

\usepackage{graphicx}
\usepackage{dcolumn}
\usepackage{bm}
\usepackage{acronym}
\usepackage{booktabs}
\usepackage{float}
\usepackage{pifont}
\usepackage{siunitx}
\usepackage{placeins}
\usepackage{mathrsfs}
\usepackage[utf8]{inputenc}
\usepackage[normalem]{ulem}
\usepackage[breaklinks,colorlinks,citecolor=blue]{hyperref}
\usepackage{graphicx,psfrag}
\usepackage{placeins}
\usepackage{multirow}
\usepackage{comment}
\usepackage{ulem}
\usepackage{hyperref}
\usepackage{color}

\newcommand{\msun}{\mathrm{\, M_\odot}}

\newcommand{\cmark}{\ding{51}}%
\newcommand{\xmark}{\ding{55}}%

\definecolor{cyan}{rgb}{0,0.9,0.9}
\definecolor{orange}{rgb}{0.9,0.5,0}
\definecolor{magenta}{rgb}{1,0,1}
\definecolor{purple}{rgb}{0.8,0.4,0.8}
\definecolor{darkgreen}{rgb}{0.0,0.5,0.0}
\definecolor{gray}{rgb}{0.8242,0.8242,0.8242}
\definecolor{cadmiumgreen}{rgb}{0.0, 0.42, 0.24}

\def\Msun{{\rm M_{\odot}}}
\def\kna{kilonova afterglow}
\def\lcs{light curves}
\def\lc{light curve}
\def\ie{\textit{i.e.}}

\def\muJy{\mu{\rm Jy}}
\def\Ek{E_{\rm ej;kin}}

\def\Fp{F_{\rm peak}}
\def\tp{t_{\rm peak}}

\begin{document}

\preprint{APS/123-QED}

\title{Signatures of deconfined quark phases in binary neutron star mergers}

\author{Aviral Prakash}
\affiliation{Institute for Gravitation \& the Cosmos, The Pennsylvania State University, University Park PA 16802, USA}
\affiliation{Department of Physics, The Pennsylvania State University, University Park PA 16802, USA}

\author{David Radice}
\affiliation{Institute for Gravitation \& the Cosmos, The Pennsylvania State University, University Park PA 16802, USA}
\affiliation{Department of Physics, The Pennsylvania State University, University Park PA 16802, USA}
\affiliation{Department of Astronomy \& Astrophysics,  The Pennsylvania State University, University Park PA 16802, USA}

\author{Domenico Logoteta}
\affiliation{Dipartimento di Fisica, Universit\`{a} di Pisa, Largo B.  Pontecorvo, 3 I-56127 Pisa, Italy}
\affiliation{INFN, Sezione di Pisa, Largo B. Pontecorvo, 3 I-56127 Pisa, Italy}

\author{Albino Perego}
\affiliation{Dipartimento di Fisica, Universit\'{a} di Trento, Via Sommarive 14, 38123 Trento, Italy}
\affiliation{INFN-TIFPA,Trento Institute for Fundamental Physics and Applications, via Sommarive 14, I-38123 Trento, Italy}

\author{Vsevolod Nedora}
\affiliation{Theoretisch-Physikalisches Institut, Friedrich-SchillerUniversit{\"a}t Jena, 07743, Jena, Germany}

\author{Ignazio Bombaci}
\affiliation{Dipartimento di Fisica, Universit\`{a} di Pisa, Largo B.  Pontecorvo, 3 I-56127 Pisa, Italy}
\affiliation{INFN, Sezione di Pisa, Largo B. Pontecorvo, 3 I-56127 Pisa, Italy}

\author{Rahul Kashyap}
\affiliation{Institute for Gravitation \& the Cosmos, The Pennsylvania State University, University Park PA 16802, USA}
\affiliation{Department of Physics, The Pennsylvania State University, University Park PA 16802, USA}

\author{Sebastiano Bernuzzi}
\affiliation{Theoretisch-Physikalisches Institut, Friedrich-SchillerUniversit{\"a}t Jena, 07743, Jena, Germany}

\author{Andrea Endrizzi}
\affiliation{Theoretisch-Physikalisches Institut, Friedrich-SchillerUniversit{\"a}t Jena, 07743, Jena, Germany}

\begin{abstract}
We investigate the quark deconfinement phase transition in the context of binary neutron star (BNS) mergers. We treat hadronic matter using a Brueckner-Hartree-Fock quantum many-body approach and modern two-body and three-body nuclear interactions derived within chiral effective field theory. Quark matter is modelled using an extended version of the bag model. We combine these approaches to construct a new finite-temperature composition-dependent equation of state (EOS) with a first order phase transition between hadrons and deconfined quarks. We perform numerical relativity simulations of BNS mergers with this new EOS and compare results obtained with or without the deconfinment phase transition. We find that deconfined quark production in a neutron star merger results from matter crossing the phase boundary over a wide range of temperatures and densities. The softening of the EOS due to the phase transition causes the merger remnants to be more compact and to collapse to a black hole (BH) at earlier times. The phase transition is imprinted on the postmerger gravitational wave (GW) signal duration, amplitude, and peak frequency. However, this imprint is only detectable for binaries with sufficiently long-lived remnants. Moreover, the phase transition does not result in significant deviations from quasi-universal relations for the postmerger GW peak frequency. Consequently, the postmerger GW peak frequency alone is not sufficient to conclusively exclude or confirm the presence of a phase transition in a BNS merger. We also study the impact of the phase transition on dynamical ejecta, remnant accretion disk masses, r-process nucleosynthetic yields and associated electromagnetic (EM) counterparts. While there are differences in the EM counterparts and nucleosynthesis yields between the purely hadronic models and the models with phase transitions, these can be primarily ascribed to the difference in remnant collapse time between the two, so they are degenerate with other effects. An exception is the non-thermal afterglow caused by the interaction of the fastest component of the dynamical ejecta and the interstellar medium, which is systematically boosted in the binaries with phase transition as a consequence of the more violent merger they experience.
\end{abstract}

\maketitle

\section{\label{sec:level1}Introduction}
Strong interacting matter, described by quantum chromodynamics (QCD), possesses a rich phase structure \cite{2009RvMP...81.1031B,2008RvMP...80.1455A,Anglani:2013gfu}. At low enough temperatures and densities the QCD phase diagram is populated by hadronic matter, i.e. a phase where quarks and gluons are confined within baryons and mesons. High precision QCD calculations on a space-time lattice at zero baryon chemical potential (i.e. zero baryon density) have clearly shown that at high temperature and for physical values of the quark masses, quarks and gluons become the most relevant degrees of freedom. The transition to this quark gluon plasma phase is a crossover \cite{2006Natur.443..675A,Bazavov:2011nk} rather than a real phase transition with a pseudo-critical temperature of about 155~MeV. The phase with deconfined quarks and gluons has been observed in heavy-ion collision experiments at very high beam energies probing the high temperature and low density region of the QCD phase diagram (see e.g. \cite{Busza:2018rrf} and the references therein).

A transition to a phase with deconfined quarks and gluons is also expected in the region with low or moderate temperatures ($T = 0 $ -- $100\,$MeV) and large densities (several times the nuclear saturation density $\rho_{nuc} \sim 2.7 \times 10^{14} \rm{g~ cm^{-3}}$).
This is the region of the QCD phase diagram that is mapped  by neutron star (NS) interiors, the hot and dense matter formed during core-collapse supernovae and BNS mergers. In fact, since a long time it has been proposed that quark matter composed of the three lightest quark flavors, namely up ($u$), down ($d$) and strange ($s$) quarks, can exist inside the core of NSs (the so called hybrid stars) or form a new type of self-bound compact stars (strange stars) which are completely made of strange quark matter (see e.g., \cite{1997csnp.book.....G}). Whether in this region of the QCD phase diagram the quark deconfinement phase transition is of the first order with a critical endpoint, or whether it proceeds smoothly through a crossover is still an open question. The latter cannot be answered by lattice QCD calculations due to the so called ``sign problem'', which makes all known lattice QCD methods at finite baryon chemical potential inapplicable. New dedicated experiments under construction at future facilities as the Compressed Baryonic Matter (CBM) experiment \cite{senger2021} at the Facility for Antiproton and Ion Research (FAIR)  will clarify this and others fundamental questions on dense QCD matter in the upcoming years.

The discovery of gravitational waves (GWs) from the BNS merger GW170817 by Advanced LIGO and Advanced VIRGO \cite{TheLIGOScientific:2017qsa}, complemented by the subsequent observations of electromagnetic (EM) counterparts by a host of earth and space-based telescopes \cite{GBM:2017lvd}, has ushered in the new field of multimessenger astronomy with GWs. It is now possible to indirectly probe the nature of the dense and hot matter created in BNS mergers through multimessenger observations. Numerical simulations with sophisticated multiphysics are required to model the highly dynamical post-merger evolution of BNS systems and bridge the gap between the fundamental physics of mergers and observational data.

The works by Most et al.~\cite{Most:2018eaw, Most:2019onn}, Bauswein et al.~\cite{Bauswein:2018bma, Bauswein:2020ggy}, Weigh et al.~\cite{Weih:2019xvw}, Liebling et al.~\cite{Liebling:2020dhf}, and Blacker et al.~\cite{Blacker:2020nlq} extensively studied QCD phase transitions in BNS mergers by contrasting simulation results obtained with EOS models in which the  QCD phase transition was included or excluded. Most et al.~\cite{Most:2018eaw, Most:2019onn} employed a chiral mean field model with a first order phase transition from hadrons to quarks, which also included hyperons. They found that, for their particular choice of EOS, a first-order QCD phase transition induced similar qualitative differences in the postmerger dynamics and the associated GW signal as the appearance of hyperons \cite{Sekiguchi:2011mc, Radice:2016rys}. In particular, the appearance of quarks was rapidly followed by BH formation in their studies. They also identified a small dephasing in the postmerger GW signal, which was claimed to be a unique signature of the formation of quarks. However, it is not clear that such a dephasing is significant given the numerical uncertainties in their simulations.

The studies of Bauswein et al.~\cite{Bauswein:2018bma, Bauswein:2020ggy} differed from the previous ones in several aspects. They employed an EOS that contained an extended mixed phase of quarks and hadrons \cite{Fischer:2017lag}, while the EOS adopted by Most et al.~\cite{Most:2018eaw, Most:2019onn} had a rapid transition to pure quark matter. Bauswein et al.~\cite{Bauswein:2018bma, Bauswein:2020ggy} also used a smoothed particle hydrodanamics (SPH) code instead of a grid based code and employed the conformally flat approximation to general relativity (GR). The simulations of Bauswein et al.~\cite{Bauswein:2018bma, Bauswein:2020ggy} resulted in the formation of quadrupolarly deformed hybrid remnants with hadronic envelopes and deconfined quarks in their cores that did not immediately collapse to BHs. These remnants were found to emit GWs at a substantially higher frequency than their hadronic counterparts. In particular, the GW signal from hybrid remnants violated empirical relations between certain properties of the binaries that can be measured from the inspiral signal and their postmerger peak frequencies \cite{Bauswein:2011tp, Hotokezaka:2013iia, Bernuzzi:2014kca, Rezzolla:2016nxn, Zappa:2017xba}. Since these relations are known to hold for all hadronic EOSs \cite{Breschi:2019srl}, the detection of a signal violating them would be a smoking gun evidence for the presence of a first order phase transition.

The apparent discrepancies between the results of Most et  al.~\cite{Most:2018eaw, Most:2019onn} and Bauswein et al.~\cite{Bauswein:2018bma, Bauswein:2020ggy} have been addressed by Weih et al. \cite{Weih:2019xvw}. This study considered a purely phenomenological description of the EOS using a piecewise polytropic ansatz. Weih et al. \cite{Weih:2019xvw} found that, depending on the characteristic of the EOS and of the phase transition, there were different possibilities. A shift in the postmerger peak GW frequency was found to occur only for models in which the phase transition is not immediately followed by BH formation. Moreover, in the cases in which the phase transition was delayed from the onset of the merger it was possible for the postmerger GW spectrum to display two peaks: one associated with the hadronic remnant prior to the phase transition, and one associated with the hybrid remnant after the phase transition. More recently, Liebling et al.~\cite{Liebling:2020dhf} used the same phenomenological ansatz as Weih et al. \cite{Weih:2019xvw}. They confirmed the previous findings and also studied the impact of phase transitions on the topology of the magnetic field of the stars.

A recent work by Blacker et al. \cite{Blacker:2020nlq} attempted to derive a methodology to constrain the onset density of a deconfinement phase transition in BNS mergers. They investigated the effects of quark deconfinement over a substantial range of NS masses. They used EOS framework as Refs.~\cite{Bauswein:2018bma, Bauswein:2020ggy}, but varied some of the model parameters. They claimed that with several measurements of the postmerger peak frequency for different binary masses it would be possible to constrain the density threshold for quark deconfinement at zero temperature.

The related scenario of the merger of self bound compact stars or strange stars was considered by Bauswein et al.~\cite{Bauswein:2008gx, Bauswein:2009im} and Zhu et al.~\cite{Zhu:2021xlu}, while the merger between hadronic NSs and strange quark stars was considered in De Pietri et al.~\cite{DePietri:2019khb}. These studies highlighted some potential GW and EM signatures for strange quark stars. Such scenarios could be independently constrained from upper bounds on the flux of strangelets generated in such mergers \cite{Madsen:1989pg, Bauswein:2008gx}.

A concordant picture has started to emerge on the possible role of QCD phase transition in mergers. However, there are still many open questions. What are the prospects for constraining a phase transition given a realistic BNS population? How generic are the signatures identified by Bauswein et al.~\cite{Bauswein:2008gx, Bauswein:2009im} and Blacker et al. \cite{Blacker:2020nlq}? Can EM counterparts and nucleosynthesis yield provide an independent constraint? In this work, we begin to address these questions using BNS merger simulations in full general relativity. We use a state of the art microphysical nuclear EOS for the hadronic phase and a phenomenological bag model EOS for the quark phase, coupled to a model for neutrino transport. We focus on a wider range of total binary mass and mass ratios than what has been considered in the past and study, for the first time, the possible signature of phase transitions in kilonovae, r-process nucleosynthesis yield, and afterglows of BNS mergers. We confirm that QCD phase transitions could leave a detectable imprint on the postmerger GW signal. However, such signature might not be easily identifiable. Indeed, the differences between our hadronic and mixed quark binaries are of the same order as differences between different hadronic models already presented in the literature. Bauswein et al. \cite{Bauswein:2019skm} also investigated the mass ejection rates from BNS mergers in the context of EOSs with deconfined quarks and reported on the absence of characteristic signatures resulting from the quark deconfinement. Likewise, we do not find any smoking gun signature of a phase transition in the kilonova or nucleosynthesis yields, but we find that the onset of a QCD phase transition can lead to more energetic bounces of the remnant. These, in turn, result in the ejection of a small amount of material to trans-relativistic velocities which could power particularly bright non-thermal afterglows. However, this effect cannot be presently used to constrain phase transition in mergers owing to the large uncertainties in the physics of the shock launched by the ejecta in the interstellar medium (ISM).

The paper is organized as follows. In the subsequent sections \ref{sec:equations_of_state} and \ref{sec:numerical_Setup}, we describe, respectively, the details of the EOSs and the numerical infrastructure for the calculations presented in the rest of the paper. In section \ref{sec:merger_dynamics}, we describe the dynamics of the merger. In particular, in sub-section \ref{subsec:qualitative_dynamics} we comment upon the qualitative features of the evolution of a BNS merger with a QCD phase transition and in sub-section \ref{subsec:dynamics_of_the_phase_transition}, we probe the thermodynamic properties of the matter produced in mergers using Lagrangian tracer particles. Section \ref{sec:gravitatoinal_waves} is devoted to the study of the GW signatures of such a transition. A discussion about the properties of the outflow from such mergers and accretion disks surrounding the remnant follows in section \ref{sec:dynamical_ejecta_and_disks}. Section \ref{sec:em_signatures} is dedicated to the discussion of possible EM signatures coming from mergers exhibiting a QCD phase transition. In particular, we compute the kilonova lightcurves at early times after the merger and the late-time afterglow in sub-sections \ref{subsec:kilonova_light_curves} and \ref{subsec:kilonova_afterglow} respectively. We finally culminate the paper with conclusions in section \ref{sec:conclusions}.


\section{Equations of State}
\label{sec:equations_of_state}

\subsection{Hadronic matter}
In the first case (hadronic matter) we assume that the dense and hot matter formed during BNS mergers can be modeled, up to the relevant temperatures and densities, as a uniform electric-charge-neutral fluid of neutrons, protons, electrons, positrons, and photons. Recently a  new microscopic EOS for this system has been obtained in Bombaci et al.~\cite{Bombaci:2018ksa} (hereafter the BL EOS) for the zero temperature case, using the Brueckner-Hartree-Fock (BHF) quantum many-body approach (see \cite{Bombaci:2018ksa} and references therein) starting from modern two-body and three-body nuclear interactions derived within chiral effective field theory (ChEFT) (e.g. \cite{2011PhR...503....1M, 2020RvMP...92b5004H}). These chiral nuclear interactions reproduce with high accuracy the nucleon-nucleon (NN) scattering data and the experimental binding energies of light (A = 3, 4) atomic nuclei. The BL EOS reproduces the empirical properties (energy per nucleon, symmetry energy and its slope parameter $L$, incompressibility) of nuclear matter at saturation density ($n_0 = 0.16\,\mathrm{fm}^{-3}$; see Logoteta et al.~\cite{Logoteta:2016nzc}), it does not violate causality (i.e. $v_s < c$, with $v_s$ being the speed of sound in the nuclear medium), and it is consistent (see figure\ 2 in \cite{Bombaci:2018ksa}) with the measured elliptic flow of matter in heavy-ion collisions experiments \cite{Danielewicz:2002pu}.

\begin{figure}
\includegraphics[width=\columnwidth]{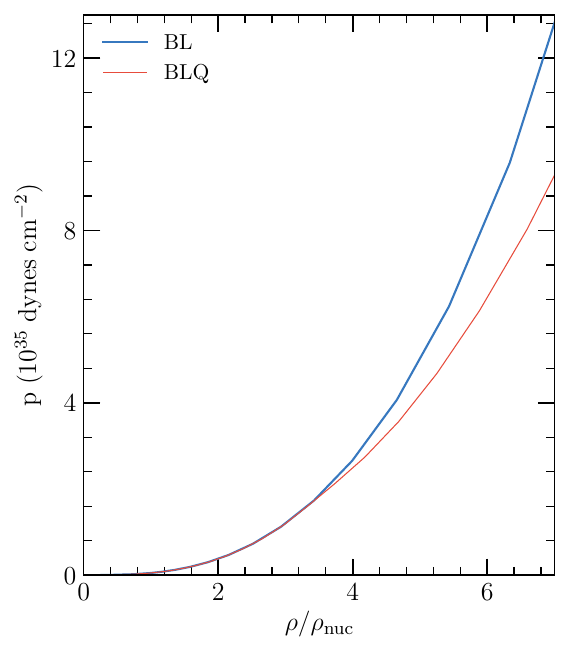}
\includegraphics[width=\columnwidth]{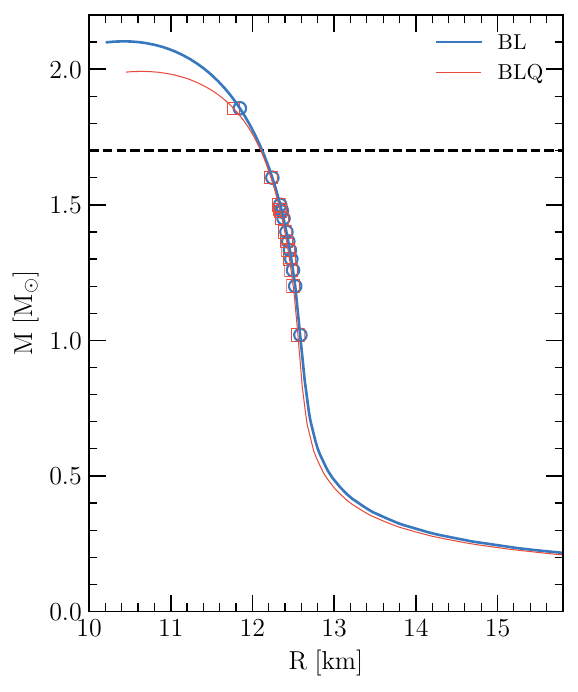}
\caption{The pressure-density variation at $T=0$ and the mass-radius relationship for isolated, cold ($T=0$), $\beta$-equilibrated, and spherically symmetric neutron stars constructed with the two equations of state used in this work. The circle and square markers represent the individual masses of the neutron stars simulated for BL and BLQ EOS respectively. The BLQ mass-radius sequence departs from the BL sequence for neutron stars having a mass $\rm{M} \gtrsim 1.7\, \msun$. These stars possess in fact a core made of hadron-quark mixed matter.}
\label{fig:EOS}
\end{figure}

When computing static neutron star configurations, the BL EOS (for the $\beta$-stable case) gives:
(i) a maximum mass $\rm{M}_{\rm max} = 2.08\,\msun$ and a corresponding radius $\rm{R}(M_{\rm max}) = 10.22\,\mathrm{km}$;
(ii) a quadrupolar tidal polarizability coefficient $\Lambda_{1.4} = 385$ (for the $1.4\,\msun$ neutron star \cite{Logoteta:2019cyb}) compatible with the constraints derived from GW170817 \cite{TheLIGOScientific:2017qsa}; and 
(iii) a threshold mass for the prompt collapse of a $q=1$ BNS system to BH as $\rm{M}_{\mathrm{threshold}} = 2.925\, \msun$ (Section~\ref{subsec:qualitative_dynamics}) indicating that GW170817 is compatible with being a NS-NS system if NSs are described by this EOS. With the addition of a thermal contribution, using the so-called $\Gamma$-law, the BL EOS has been used in BNS merger simulations by Endrizzi et al.~\cite{Endrizzi:2018uwl}.

Very recently, the BL EOS has been extended to finite temperature and to arbitrary proton fractions~\cite{Logoteta:2020yxf}. In the following we will refer to this finite-temperature EOS model as the BLh EOS. The BLh EOS has been applied in Bernuzzi et al.~\cite{Bernuzzi:2020txg, Vsevolod:2020pak} to asymmetric neutron star mergers $(q \neq 1)$ with chirp mass $1.188\, \msun$ corresponding to the measured one in the case of GW170817.

\subsection{Deconfined quark matter}
The second EOS model (hereafter the BLQ EOS) used in our work describes the thermodynamical properties of hadronic-quark hybrid matter. We assume that at high enough temperatures and densities reached during a BNS merger, stellar matter undergoes a transition to a phase with deconfined quarks (quark matter) and in addition we assume this phase transition to be of the first-order. Quark matter could also be present in the NS cores prior to merger in the case of sufficiently massive NSs (see figure \ref{fig:EOS}) and for particular choices of the quark matter EOS parameters. To describe the hadronic phase (i.e. quarks confined in neutrons and protons in our case) of hybrid matter we make use of the BLh EOS described above.

To model the quark phase we use an extended version of the phenomenological bag model EOS which includes the effects of gluon mediated QCD interactions between quarks up to the second order in the QCD coupling $\alpha_s$ \cite{2001PhRvD..63l1702F, 2005ApJ...629..969A, 2011ApJ...740L..14W}. The grand canonical potential per unit volume can be written as (we use units where $\hbar = c = 1$):
\begin{equation}
\label{eos}
\Omega = \sum_{i=u,d,s,e} \Omega_i^0 +  \frac{3}{4\pi^2}(1 - a_4)
\mu^4 + B_{\rm eff}\, ,
\end{equation}
where $\Omega_i^0$ is the grand canonical potential density for {\it u}, {\it d}, {\it s} quarks and electrons (and their antiparticles), described as ideal relativistic Fermi gases.
The second term on the right hand side of Eq.(\ref{eos})
accounts for the perturbative QCD corrections to $\mathcal{O}(\alpha_s^2)$
\cite{2001PhRvD..63l1702F,2005ApJ...629..969A,2011ApJ...740L..14W} and its value represents the degree of deviations from
an ideal gas EOS, with $a_4 = 1$ corresponding to the ideal case.
The chemical potential $\mu$ in Eq.~(\ref{eos}) can be written in terms of the {\it u}, {\it d} and {\it s} quark chemical potentials as $\mu= (\mu_u + \mu_d + \mu_s)/3$.
The term $B_{\rm eff}$ is an effective bag constant which takes into accounts in a phenomenological way nonperturbative aspects of QCD.

At finite temperature, the ideal gas contributions to $\Omega$ provided by fermions and antifermions can be calculated by computing the corresponding Fermi integrals for a given temperature $T$ and chemical potential $\mu_i$  (see e.g. \cite{1986bhwd.book.....S}):
\begin{eqnarray}
\Omega^0_{i}(T,\mu_i)&=& - \frac{1}{3} \frac{g_i}{2 \pi^2} \int_0^\infty
k^2 dk\,\,k\, v
\nonumber \\
&\times& \left[f (k,\mu_i) + f (k,-\mu_i) \right]
\label{Omega_0}
\end{eqnarray}   
where $v = k/E_i$ is the particle velocity (with $E_i(k)=(k^2 + m_i^2)^{1/2}$),
and $f(k,\pm \mu_i)$ are the Fermi distribution functions with chemical
potentials for particles ($+ \mu_i$) and antiparticles ($- \mu_i$):
\begin{eqnarray}
 f(k, \pm \mu_i) = \frac{1}{e^{(E_i(k) \mp \mu_i)/T} +1},
\end{eqnarray}
the degeneracy factor is $g_i = 2$ for electrons and  $g_i = 6$ for each quark flavor. 
We neglect the temperature dependence of the last two terms in Eq.(\ref{eos}). 

The total entropy density
\begin{equation}
\label{tot_entropy}
  s = \sum_{i=u,d,s,e} s_i\,
\end{equation}
can be calculated using the ideal Fermi gas approximation for each fermionic particle species \cite{Fetter-Walecka:book1971}:
\begin{eqnarray}
s_i (T,\mu_i) &=& - \frac{g_i}{2\pi^2} \int_0^\infty k^2 dk
\left[ f (k,\mu_i) \mathrm{ln} f (k,\mu_i) \right.
\nonumber \\
&+& ( 1 - f(k,\mu_i)) \mathrm{ln}(1-f(k,\mu_i))
\nonumber\\
&+&
f(k,-\mu_i) \mathrm{ln} f(k,-\mu_i)
\nonumber \\
&+& \left. ( 1 - f (k,-\mu_i)) \mathrm{ln}(1-f (k,-\mu_i))
\right]
\label{entropy_f}
\end{eqnarray}

Using standard thermodynamical relations, the energy density can be written as:
\begin{equation}
\label{endens}
\varepsilon = \Omega + \sum_{i=u,d,s,e}{\mu_i n_i} + Ts ,
\end{equation}
where $n_i$ is the number density for each particle species which can be calculated as:
\begin{equation}
\label{numdens}
      n_i = - \bigg(\frac{\partial\Omega}{\partial \mu_i}\bigg)_{T,V}
\end{equation}
and the total baryon number density is:
\begin{equation}
\label{n_B}
       n_B =  \frac{1}{3}(n_u + n_d + n_s)\,.
\end{equation}

We next assume a first order hadron-quark phase transition and, following Glendenning \cite{Glendenning:1992vb}, we require global electric charge neutrality of bulk stellar matter. An important consequence of imposing global charge neutrality is that the hadronic and the quark phases can coexist for a finite range of pressures. This treatment of the phase transition is known in the literature as the Gibbs construction for the hadron-quark mixed phase. In this case the Gibbs conditions for phase equilibrium can be written  as  \cite{Glendenning:1992vb}:
\begin{eqnarray}
\mu_{b,H} & = & \mu_{b,Q} \equiv \mu_b \, , \\
\mu_{q,H} & = & \mu_{q,Q} \equiv \mu_q \, , \\
\label{g1}
T_H & = & T_Q \equiv T \, , \\
\label{g2}
P_H(\mu_b, \mu_q, T) & = & P_Q(\mu_b, \mu_q, T) \, ,
\end{eqnarray}
where the subscript $H$ and $Q$ refer to physical quantities in the hadronic and in the quark phase respectively, while the baryon chemical potential, $\mu_b$, and the electric chemical potential, $\mu_q$, are two independent chemical potentials corresponding respectively to the global conservation of the baryon number and the electric charge.
In the pure hadronic phase $\mu_b = \mu_n$, the neutron chemical potential, and $\mu_q = \mu_e$, the electron chemical potential.
In the quark phase the quark chemical potentials can be written as:
\begin{eqnarray}
\mu_u = \frac{1}{3}(\mu_b - 2 \mu_q)  = \frac{2}{3} \mu_p -\frac{1}{3} \mu_n  \, ,\\
\mu_d = \frac{1}{3}(\mu_b + \mu_q) = \frac{2}{3} \mu_n -\frac{1}{3} \mu_p \;.
\end{eqnarray}
Weak reactions of the type:
\begin{eqnarray}
  d + u \leftrightarrow u + s\,  \\
  u + e^-  \leftrightarrow s + \nu_e
\end{eqnarray}
will change the strangeness content of the just deconfined quark matter \cite{Bombaci:2016xuj} to minimize the energy per baryon of the system.
Since the typical time-scale for weak interaction processes $t_{w} \lesssim (10^{-8}$--$10^{-10})\,\mathrm{s}$ is significantly shorter than the hydrodynamics timescales inside the remnant, we neglect detailed reaction rates involving quarks and neutrinos, and we consider $\beta$-stable strange quark matter with the strange quark chemical potential $\mu_s = \mu_d$.

In the present work we take $m_u = m_d = 0$, $m_s = 100~\rm{MeV}$,
$B_{\rm eff}^{1/4} = 180\,\mathrm{MeV}$ and $a_4 = 0.4$. With these values of the EOS parameters for the quark phase, and with the BL EOS for the hadronic phase, we obtain the NS mass-radius curves shown in figure~\ref{fig:EOS}. In particular, we find identical radii and NS structure for BL and BLQ when considering stars with masses up to $\rm{M} \simeq 1.7\,\msun$. That is, up to the onset of the phase transition. After the onset of the phase transition, the BLQ EOS becomes less stiff than BL and predicts more compact NSs and a lower maximum mass of $\rm{M}_{\rm max} = 1.99\,\msun$. The corresponding radius is $\rm{R}(\rm{M}_{\max}) = 10.46\ {\rm km}$. The threshold for prompt BH formation for a $q = 1$ BNS merger with BLQ EOS is found to be $2.825\,\msun$ (see Section\;\ref{subsec:qualitative_dynamics}).


\section{Numerical Setup}
\label{sec:numerical_Setup}
We evolve our systems in full general relativity along with high order convergence schemes for general relativistic hydrodynamics using the $\tt{WhiskyTHC}$ code \cite{Radice:2012cu, Radice:2013hxh, Radice:2013xpa, Radice:2016dwd, Radice:2018pdn}. The spacetime metric is evolved using the Z4c formulation \cite{Bernuzzi:2009ex, Hilditch:2012fp} of the Einstein's equations implemented in the $\tt{CTGamma}$ thorn \cite{Pollney:2009yz, Reisswig:2013sqa} of $\tt{Einstein Toolkit}$ \cite{Loffler:2011ay}. Our simulations make use of the \texttt{Carpet} adaptive mesh refinement (AMR) framework \cite{Schnetter:2003rb, Reisswig:2012nc}, which implements the Berger-Oliger scheme with refluxing \cite{Berger:1984zza, 1989JCoPh..82...64B}. For the relativistic hydrodynamics, $\tt{WhiskyTHC}$ evolves proton number and neutron number densities separately to ensure detailed conservation given by:
\begin{equation}
    \nabla_{\mu}(n_{p,n}\;u^{\mu}) = R_{p, n}
\end{equation}
where $n_{p, n}$ are the number densities of (free or bound) protons and neutrons respectively, $u^{\mu}$ is the fluid four-velocity and $R_{p, n}$ are the net lepton number deposition rates due to the emission and absorption of electron neutrinos and anti-neutrinos. Due to charge neutrality, the relative amount of neutrons and protons is expressed in terms of ${\rm Y}_e$, i.e. the electron fraction given by $n_p/(n_p+n_n)$. The evolution of the energy-momentum tensor takes the following form:
\begin{equation}
    \nabla_{\mu}T^{\mu\nu} = Qu^{\nu} \, ,
    \label{tmunu}
\end{equation}
where $Q$ is the net energy deposition rate due to the emission and absorption of neutrinos and anti neutrinos of all flavours. We consider the relativistic fluid of the neutron star to be a perfect fluid with no shears, viscosity, or heat conduction, that is:
\begin{equation}
    T^{\mu\nu} = (e+p) u^{\mu}u^{\nu} + pg^{\mu\nu}.
\end{equation}
Here $e$ is the total energy density, $p$ the isotropic pressure and $g^{\mu\nu}$ the spacetime metric. For additional details on the numerical schemes used to discretize the above equations and the specifics of the neutrino leakage scheme, we refer to Radice et al.~\cite{Radice:2018pdn} and the references therein.

In order to record the thermodynamic history of the relativistic flow in and around the remnant's core, we inject Lagrange tracer particles in the fluid frame. These are fiducial particles that are advected with the flow according to:
\begin{equation}
    \frac{{\rm d}x^i}{{\rm d}t} = \alpha v^{i} - \beta^{i}.
\end{equation}
Here $\alpha$ is the lapse function, $v^i$ is the three-velocity of the fluid, and $\beta^i$ is the shift vector.

\begin{table} 
\caption{ 
 A summary of the properties of non-spinning isolated NSs used for constructing the initial data with the BL EOS. $\rm{M_1}$ represents the primary (heavier) mass in the binary and $\rm{M_2}$ is the secondary mass. $\rm{M}$ represents the total mass whereas q and $\nu$ represent the mass ratio and the symmetric mass ratio of the binary. $\Lambda_i$s $(i \in \{1, 2\})$ are the respective quadrupolar tidal polarizability coefficients of the individual stars and $\widetilde{\Lambda}$ is a parameter defined in Eq(5) of \cite{Wade:2014vqa}. $\xi$ is a tidal parameter constructed in \cite{Breschi:2019srl} from $\Lambda_i$s. 
} 
\label{tab:inspiral_property} 
 \begin{center} {
 \begin{tabular}{c c c c c c c c c} 
 \hline\hline 
 \\ 
 $\rm{M_1}$ & $\rm{M_2}$ & $\rm{M}$ & q & $\nu $ & $\Lambda_1$ & $\Lambda_2$ & $\widetilde{\Lambda}$ & $\xi$ \\ 
 $[\rm{M}_{\odot}]$ & $[\rm{M}_{\odot}]$ & $[\rm{M}_{\odot}]$ & & & & & & \\ 
 \hline 
 \hline 
1.3 & 1.3 & 2.6 & 1.0 & 0.25 & 696 & 696 & 696 & 130 \\
1.3325 & 1.3325 & 2.67 & 1.0 & 0.25 & 595 & 595 & 595 & 111 \\
1.365 & 1.365 & 2.73 & 1.0 & 0.25 & 510 & 510 & 510 & 95 \\
1.4 & 1.4 & 2.8 & 1.0 & 0.25 & 432 & 432 & 432 & 81 \\
1.45 & 1.45 & 2.9 & 1.0 & 0.25 & 341 & 341 & 341 & 63 \\
1.475 & 1.475 & 2.95 & 1.0 & 0.25 & 303 & 303 & 303 & 56 \\
1.5 & 1.5 & 3.0 & 1.0 & 0.25 & 269 & 269 & 269 & 50 \\
1.6 & 1.6 & 3.2 & 1.0 & 0.25 & 168 & 168 & 168 & 31 \\
1.4 & 1.2 & 2.6 & 1.17 & 0.25 & 432 & 1137 & 711 & 133 \\
1.482 & 1.259 & 2.74 & 1.18 & 0.25 & 293 & 849 & 510 & 95 \\
1.856 & 1.02 & 2.88 & 1.82 & 0.23 & 46 & 2896 & 505 & 92 \\
\hline 
\hline 
\end{tabular} 
} 
\end{center} 
\end{table} 

\begin{table*}[t] 
\begin{center} 
\caption{ 
 A summary of the postmerger GW properties from all our simulations at 2 spatial resolutions. $t_{\rm{merg}}$ is the time of merger, $t_{\rm{BH}}$ is the time after merger when the system collapses to a black hole, $t_{\rm{coll}}$ is the time when the gravitational radiation from the 2,2 mode shuts down and $t_{\rm{end}}$ is the final time of the simulation. $f_{2}$ represents the dominant postmerger peak frequency of the 2,2 mode, $\Delta f_{2}$ represents the difference between the postmerger peak frequencies from the 2 EOSs and $\Delta\rm{FT}$ represents the numerical uncertainty in the Fourier transform. The binary labelled with BLh* is with GRLES (general-relativistic large-eddy simulation) and simulated using the calibrated turbulence model in \cite{Radice:2020ids}. 
} 
\label{tab:GW_property} 
\scalebox{1.05}{ 
 \begin{tabular}{c c c c c c c c c c c c c } 
 \hline \hline 
 \\[-0.9em] 
 EOS & $\rm{M_1}$ & $\rm{M_2}$ & M & q & Resolution & $t_{\text{BH}} - t_{\text{merg}}$ & $t_{\text{coll}} - t_{\text{merg}}$ & $t_{\text{end}} - t_{\text{merg}}$ & $f_{\text{2}}$  & $\Delta f_{\text{2}}$ & $\Delta \rm{FT}$ & $\rm{Prompt}$  \\ 
 & [$\msun$] & [$\msun$] & [$\msun$] & & & [$\rm{ms}$] & [$\rm{ms}$] & [$\rm{ms}$] & [kHz] & [kHz] & [kHz] & $\rm{Collapse}$ \\ 
 \hline 
 \hline 
BLh & 1.3 & 1.3 & 2.6 & 1.0 & SR & $>$18.69 & $>$18.69 & 18.69 & 2.8 &   &   & \xmark \\
BLQ & 1.3 & 1.3 & 2.6 & 1.0 & SR & 15.95 & 16.6 & 19.21 & 2.92 & 0.12 & 0.06 & \xmark \\
\hline 
BLh & 1.3325 & 1.3325 & 2.67 & 1.0 & SR & $>$36.23 & $>$36.23 & 36.23 & 2.91 &   &   & \xmark \\
BLQ & 1.3325 & 1.3325 & 2.67 & 1.0 & SR & 7.44 & 8.17 & 10.55 & 3.11 & 0.19 & 0.12 & \xmark \\
\hline 
BLh & 1.365 & 1.365 & 2.73 & 1.0 & SR & $>$101.2 & $>$101.2 & 101.2 & 3.06 &   &   & \xmark \\
BLh* & 1.365 & 1.365 & 2.73 & 1.0 & SR & $>$12.34 & $>$12.34 & 12.34 & 3.05 &   &   & \xmark \\
BLQ & 1.365 & 1.365 & 2.73 & 1.0 & SR & 4.1 & 4.82 & 12.15 & 3.18 & 0.12 & 0.21 & \xmark \\
\hline 
BLh & 1.4 & 1.4 & 2.8 & 1.0 & SR & $>$38.57 & $>$38.57 & 38.57 & 3.2 &   &   & \xmark \\
BLQ & 1.4 & 1.4 & 2.8 & 1.0 & SR & 1.92 & 2.85 & 11.6 & \xmark & \xmark & \xmark & \xmark \\
\hline 
BLh & 1.45 & 1.45 & 2.9 & 1.0 & SR & $>$14.4 & $>$14.4 & 14.4 & \xmark &   &   & \xmark \\
BLQ & 1.45 & 1.45 & 2.9 & 1.0 & SR & 0.67 & 1.68 & 4.39 & \xmark & \xmark & \xmark & \cmark \\
\hline 
BLh & 1.475 & 1.475 & 2.95 & 1.0 & SR & 0.69 & 1.93 & 11.76 & \xmark &   &   & \cmark \\
BLQ & 1.475 & 1.475 & 2.95 & 1.0 & SR & 0.57 & 1.61 & 5.66 & \xmark & \xmark & \xmark & \cmark \\
\hline 
BLh & 1.5 & 1.5 & 3.0 & 1.0 & SR & 0.56 & 1.69 & 4.76 & \xmark &   &   & \cmark \\
BLQ & 1.5 & 1.5 & 3.0 & 1.0 & SR & 0.51 & 1.54 & 3.63 & \xmark & \xmark & \xmark & \cmark \\
\hline 
BLh & 1.6 & 1.6 & 3.2 & 1.0 & SR & 0.39 & 1.32 & 3.66 & \xmark &   &   & \cmark \\
BLQ & 1.6 & 1.6 & 3.2 & 1.0 & SR & 0.39 & 1.29 & 3.64 & \xmark & \xmark & \xmark & \cmark \\
\hline 
BLh & 1.4 & 1.2 & 2.6 & 1.17 & SR & $>$105.0 & $>$105.0 & 105.0 & 2.75 &   &   & \xmark \\
BLQ & 1.4 & 1.2 & 2.6 & 1.17 & SR & 17.2 & 17.63 & 23.95 & 2.96 & 0.21 & 0.06 & \xmark \\
\hline 
BLh & 1.482 & 1.259 & 2.74 & 1.18 & SR & $>$21.59 & $>$21.59 & 21.59 & 2.97 &   &   & \xmark \\
BLQ & 1.482 & 1.259 & 2.74 & 1.18 & SR & 3.54 & 4.42 & 9.03 & 3.2 & 0.23 & 0.23 & \xmark \\
\hline 
BLh & 1.856 & 1.02 & 2.88 & 1.82 & SR & 1.02 & 1.61 & 8.31 & \xmark &   &   & \cmark \\
BLQ & 1.856 & 1.02 & 2.88 & 1.82 & SR & 0.65 & 1.74 & 8.74 & \xmark & \xmark & \xmark & \cmark \\
\hline 
\hline 
BLh & 1.3 & 1.3 & 2.6 & 1.0 & LR & $>$31.9 & $>$31.9 & 31.9 & 2.82 &   &   & \xmark \\
BLQ & 1.3 & 1.3 & 2.6 & 1.0 & LR & 12.28 & 12.81 & 19.66 & 2.94 & 0.13 & 0.08 & \xmark \\
\hline 
BLh & 1.3325 & 1.3325 & 2.67 & 1.0 & LR & $>$26.35 & $>$26.35 & 26.35 & 2.88 &   &   & \xmark \\
BLQ & 1.3325 & 1.3325 & 2.67 & 1.0 & LR & 13.19 & 13.82 & 18.65 & 3.06 & 0.19 & 0.07 & \xmark \\
\hline 
BLh & 1.365 & 1.365 & 2.73 & 1.0 & LR & $>$21.39 & $>$21.39 & 21.39 & 3.03 &   &   & \xmark \\
BLQ & 1.365 & 1.365 & 2.73 & 1.0 & LR & 4.84 & 5.44 & 8.59 & 3.06 & 0.04 & 0.18 & \xmark \\
\hline 
BLh & 1.4 & 1.4 & 2.8 & 1.0 & LR & $>$23.63 & $>$23.63 & 23.63 & 3.15 &   &   & \xmark \\
BLQ & 1.4 & 1.4 & 2.8 & 1.0 & LR & 1.91 & 2.74 & 8.02 & \xmark & \xmark & \xmark & \xmark \\
\hline 
BLh & 1.45 & 1.45 & 2.9 & 1.0 & LR & 1.85 & 2.93 & 13.3 & \xmark &   &   & \xmark \\
BLQ & 1.45 & 1.45 & 2.9 & 1.0 & LR & 0.67 & 1.53 & 9.44 & \xmark & \xmark & \xmark & \cmark \\
\hline 
BLh & 1.475 & 1.475 & 2.95 & 1.0 & LR & 0.66 & 1.73 & 4.98 & \xmark &   &   & \cmark \\
BLQ & 1.475 & 1.475 & 2.95 & 1.0 & LR & 0.55 & 1.47 & 4.46 & \xmark & \xmark & \xmark & \cmark \\
\hline 
BLh & 1.5 & 1.5 & 3.0 & 1.0 & LR & 0.56 & 1.54 & 5.06 & \xmark &   &   & \cmark \\
BLQ & 1.5 & 1.5 & 3.0 & 1.0 & LR & 0.52 & 1.43 & 4.65 & \xmark & \xmark & \xmark & \cmark \\
\hline 
BLh & 1.6 & 1.6 & 3.2 & 1.0 & LR & 0.4 & 1.28 & 5.11 & \xmark &   &   & \cmark \\
BLQ & 1.6 & 1.6 & 3.2 & 1.0 & LR & 0.4 & 1.26 & 4.76 & \xmark & \xmark & \xmark & \cmark \\
\hline 
BLh & 1.482 & 1.259 & 2.74 & 1.18 & LR & $>$20.71 & $>$20.71 & 20.71 & 2.98 &   &   & \xmark \\
BLQ & 1.482 & 1.259 & 2.74 & 1.18 & LR & 3.63 & 4.35 & 11.66 & 3.14 & 0.16 & 0.23 & \xmark \\
\hline 
BLh & 1.856 & 1.02 & 2.88 & 1.82 & LR & 0.99 & 1.59 & 9.88 & \xmark &   &   & \cmark \\
BLQ & 1.856 & 1.02 & 2.88 & 1.82 & LR & 0.62 & 1.7 & 9.96 & \xmark & \xmark & \xmark & \cmark \\
\hline 
\hline 
\end{tabular} 
}
\end{center} 
\end{table*}

The initial data for all our simulations is constituted of irrotational binaries in quasi-circular orbit at an initial separation of 45 km. These are constructed using the $\tt{Lorene}$ code by Gourgoulhon et al. \cite{Gourgoulhon:2000nn} which provides classes to solve for a wide variety of partial differential equations using multi-domain spectral methods. We use the BL EOS to construct the initial data for all our systems, including those simulated with the BLQ EOS. We simulate 11 BNS configurations varying both in their total mass and mass ratios (table \ref{tab:inspiral_property}). These include binaries with total gravitational masses ranging from 2.6 $\msun$ to 3.2 $\msun$ and covering a mass ratio range of 1 to 1.82. Among our simulations there are three systems that are targeted to GW170817, namely $1.365\ \msun - 1.365\ \msun$, $1.482\ \msun - 1.259\ \msun$ and $1.856\ \msun - 1.020\ \msun$. Each of these binaries have a chirp mass of 1.18 $\msun$ that is compatible with the observations for GW170817 \cite{TheLIGOScientific:2017qsa}. Additionally, we also simulate a binary $1.4\ \msun - 1.2\ \msun$  that is consistent with the observations of the relativistic binary pulsar PSR J1829+2456 \cite{Champion:2004hc}.

We employ an AMR structure composed of 7 refinement levels. Of these, the three outer levels are fixed, while the inner four levels are comoving with the stars during their inspiral. The finest refinement levels covers entirely the stars during the inspiral and the centrally condensed part of the remnant after the merger. We simulate the binaries at two spatial resolutions (see table \ref{tab:GW_property}): with grid resolutions of 184.6~m (standard resolution; SR) or 246.1~m (low resolution; LR) in the finest refinement level. The binary $1.4\ \msun - 1.2\ \msun$ is only simulated at SR. The time step is determined by the Courant-Friedrichs-Lewy (CFL) coefficient which is taken to be $0.125$. This small CFL, in combination with the positivity preserving limiter of \texttt{WhiskyTHC}, guarantees the positivity of the density \cite{Radice:2013xpa}.


\section{Merger Dynamics}
\label{sec:merger_dynamics}

\subsection{Qualitative Dynamics}
\label{subsec:qualitative_dynamics}

\begin{figure*}
  \centering
  \includegraphics[width=\textwidth]{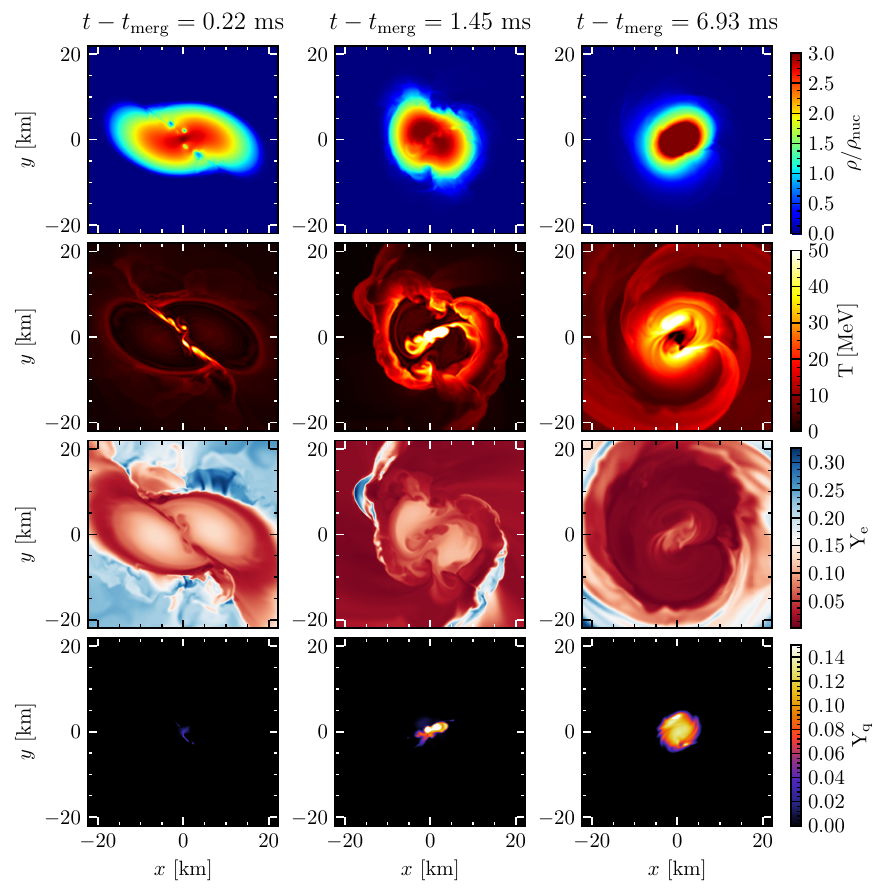}
  \caption{Evolution of the remnant's density, temperature, electron fraction and quark fraction across the xy plane for a merger of the 1.3325 $\msun$ $-$ 1.3325 $\msun$ binary. Deconfined quarks appear as matter is compressed and heated up during the merger. The quark distribution strongly correlates with the temperature distribution in the middle panel, indicating that quarks are formed due to heating during the merger. At later times, the quark distribution is centrally condensed and most strongly correlated with the density.}
  \label{fig:2D}
\end{figure*}

We start with a discussion of the qualitative dynamics of a BNS merger with a phase transition. To this aim, we show in figure~\ref{fig:2D} the postmerger evolution for the $1.3325\ \msun - 1.3325\ \msun$ binary, which is representative of our set of simulations. The figure shows the thermodynamic properties of matter in the equatorial plane. All of the binaries we have considered, with the exception of $1.856\ \msun - 1.020\ \msun$, have no deconfined quark matter during the inspiral up until merger. This is expected since, as discussed in Section \ref{sec:equations_of_state}, phase transition at zero temperature occurs only for stars more massive than about $1.7\ \msun$.

The inspiral evolutions of the BLQ and BLh binaries are identical, with the exception of  $1.856\ \msun - 1.020\ \msun$, which is discussed later. As the orbit of the binary shrinks the stars become increasingly deformed. In high mass ratio systems the deformations lead to mass transfer and the disruption of the secondary star \cite{Bernuzzi:2020txg}. In the case of comparable mass binaries there is no mass transfer between the stars up to contact.

As the stars merge, their collisional interface is heated to temperatures of up to 40 MeV \cite{Bernuzzi:2015opx, Hanauske:2016gia, Kastaun:2016yaf, Most:2018eaw, Perego:2019adq}. This hot interface is the first site of quark matter production. Over the next few milliseconds, a massive remnant begins to form with increasing densities and temperatures in and around the core. At this stage, the spatial distribution of the quark phase is largely determined by regions of high temperature and this feature is found to be a robust characteristic of all our simulations. These temperature hot-spots evolve further into a ring-like structure over the next several milliseconds which is also a feature observed in \cite{Hanauske:2016gia, Kastaun:2016yaf, Most:2018eaw}.
The hot-spots continue to dictate the spatial distribution of the quark phase, but now the extreme density in the core of the remnant, reaching up to five times nuclear saturation density, also starts to play a role in producing quark matter and determining their distribution in space. Noteworthy is the fact that both temperature and density can trigger a production of deconfined quark phase. For example, according to our EOS, for typical electron fractions $\rm{Y_e} \simeq 0.01$ and density of $3\;\rho_{\rm nuc}$, quark deconfinement can occur at temperatures as low as ${\sim}20\ {\rm MeV}$. On the other hand, at densities of the order of the nuclear saturation density, quark deconfinement requires temperatures above ${\sim}70 \ {\rm MeV}$. We explore this interplay between density and temperature for the production of quarks in more detail in subsection \ref{subsec:dynamics_of_the_phase_transition}.

\begin{figure*}
  \centering
  \includegraphics[width=0.49\textwidth]{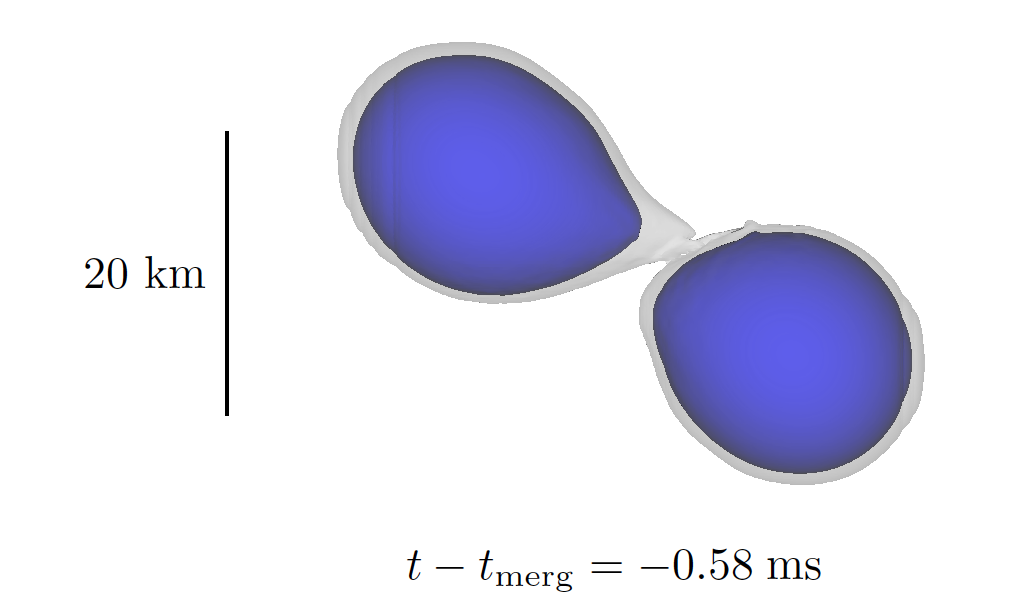}
  \includegraphics[width=0.49\textwidth]{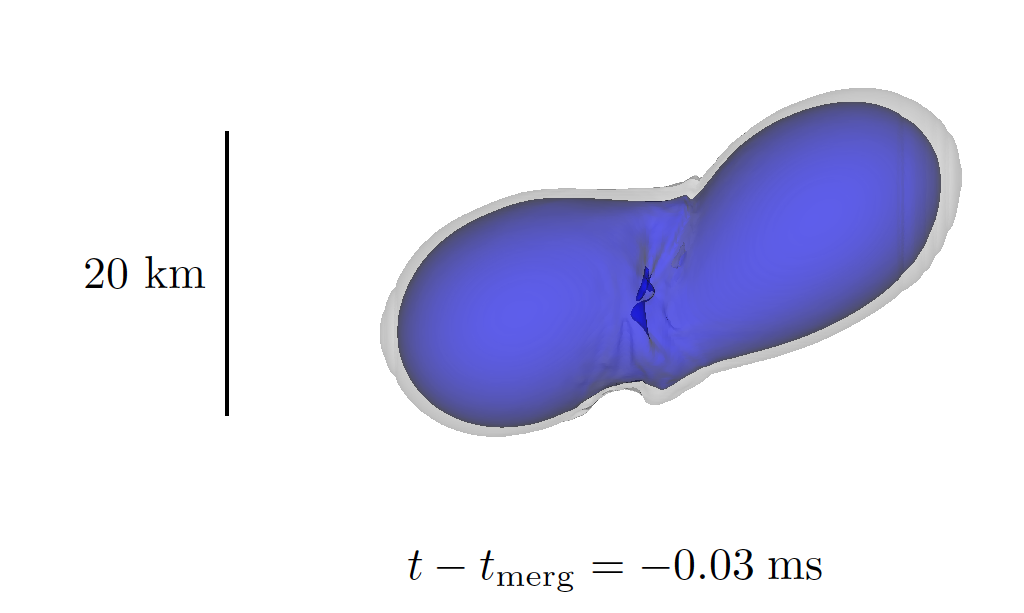}
  \includegraphics[width=0.49\textwidth]{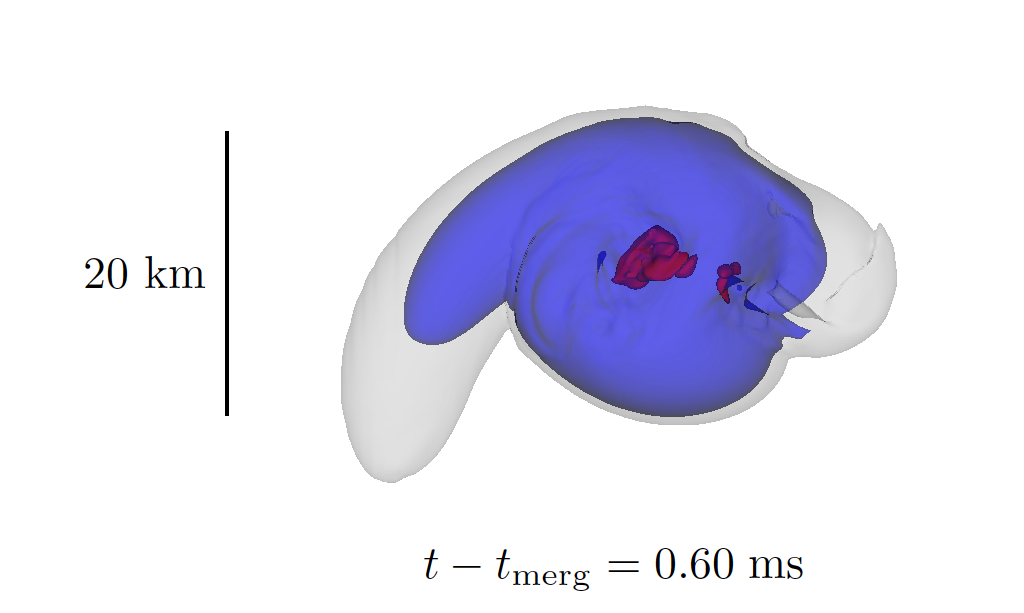}
  \includegraphics[width=0.49\textwidth]{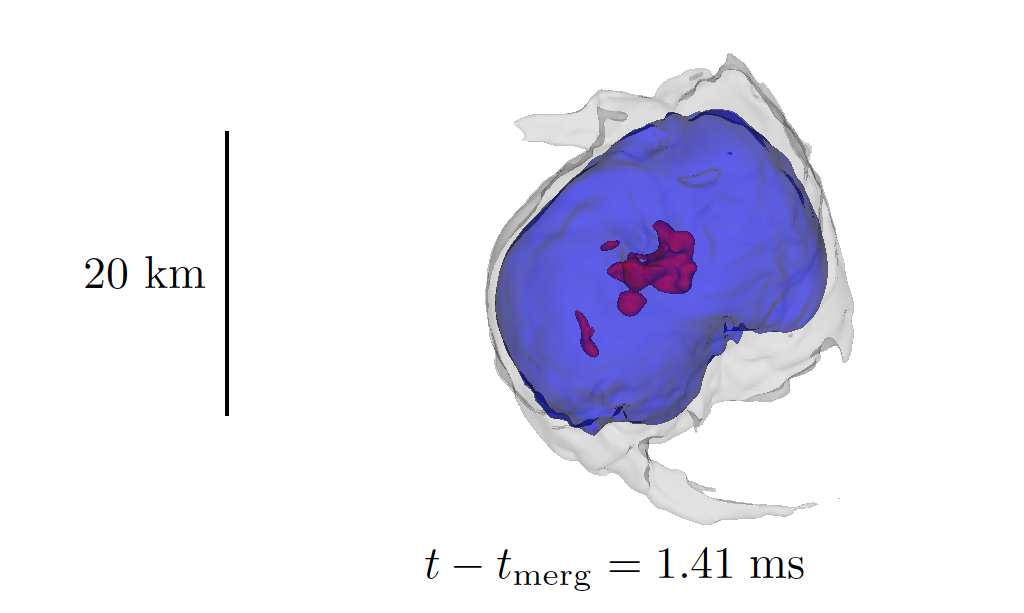}
  \caption{Evolution of a BNS merger of masses 1.4 $\msun$ and 1.2 $\msun$ evolved with the BLQ EOS. The mass configuration corresponds to the pulsar PSR J1829+2456 \cite{Champion:2004hc}. The blue and grey colour-scales represent iso-density surfaces corresponding to densities $\rm{10^{14}\;g\;cm^{-3}}$ and $\rm{10^{13}\;g\;cm^{-3}}$ respectively. The deconfined quark phase that appears near the core of the remnant after merger is represented in red.}
  \label{fig:merger3D}
\end{figure*}

As another representative example of the dynamics of the BLQ binaries, we show the evolution of the $1.4\ \msun - 1.2\ \msun$ binary in figure~\ref{fig:merger3D}. This merger proceeeds in a similar way as that of the previously discussed $1.3325\ \msun - 1.3325\ \msun$ binary. However, a qualitative difference between the dynamics of an equal and an unequal mass merger is that the quark distribution is asymmetric in the latter case. This follows from the fact that the hot-spots in unequal mass mergers are no longer spatially symmetric with respect to the remnant's centre, as also pointed out in \cite{Most:2018eaw}.

A generic feature of all of our simulations is that the remnants of the BLQ binaries are more compact and collapse earlier to a BH than the BLh binaries, for which the QCD phase transition is absent. This is because the appearance of quarks tends to soften the EOS making the NSs more compact and hence more susceptible to BH formation. Measuring the lifetimes of NS merger remnants could provide important constraints on the NS EOS. Indeed hierarchical inference studies done on collapsing neutron star models using the X-ray afterglow studies of short gamma ray bursts show evidences of a quark EOS \cite{Sarin:2020pwr}.

Binaries with mass above a critical threshold undergo prompt collapse to a BH, which we define as the absence of any bounce of the merger remnant prior to BH formation. When prompt BH formation occurs it is difficult to differentiate between systems with or without a QCD phase transition solely on the basis of postmerger GW signal, since it is effectively absent. However, the differences between the BLh and BLQ EOS manifest as a lowering of the mass threshold for the prompt BH collapse from $\rm{M} = 2.925\msun$ for BLh EOS to $\rm{M} = 2.825 \msun$ in the case of BLQ EOS (see table \ref{tab:GW_property} and Kashyap et al. 2021 in preparation). This is in agreement with the claim that phase transitions can lower the threshold towards prompt BH collapse by Bauswein et al.~\cite{Bauswein:2020ggy}.



\subsection{Dynamics of the phase transition}
\label{subsec:dynamics_of_the_phase_transition}

We follow a methodology along the lines of \cite{Most:2018eaw, Hanauske:2019qgs, Perego:2019adq, Most:2019onn} to identify the thermodynamic conditions reached in BNS mergers and whether or not these conditions are conducive towards the production of deconfined quark matter. In particular, as discussed in section \ref{sec:numerical_Setup}, we track the thermodynamic properties of the NS material in and around the core using Lagrangian tracer particles. This allows us to record the thermodynamical evolution of individual ``fluid elements''. We primarily discuss the case of the $1.3325\ \msun - 1.3325\ \msun$ binary, which is representative of most of our simulations.

The BLQ EOS provides the quark fraction $\mathrm{Y_q}$ as a function of the matter temperature $\mathrm{T}$, density $\mathrm{\rho}$, and electron fraction $\mathrm{Y_e}$, i.e. $\mathrm{Y_q = Y_q(\rho, T, Y_e)}$. As the dynamics of the phase transition takes place in and around the core of the remnant, where $\mathrm{Y_e}$ does not exceed 0.15 (see figure \ref{fig:2D}), it is reasonable to approximate the full phase diagram of the EOS with a two dimensional analog obtained by averaging $\mathrm{Y_e}$ over the range $\rm{Y_e}\in[0.0, 0.15]$. We remark that this is done for illustrative purposes only and that no such approximation is made in the simulations. There are no qualitative differences between the $\rm{Y_e}$-averaged $\langle\rm{Y_q}\rangle$ and $\rm{Y_q}$ in the range of $\rm{Y_e}$'s considered here.
%
%

\begin{figure*}[t]
  \centering
  \includegraphics[width=0.49\textwidth]{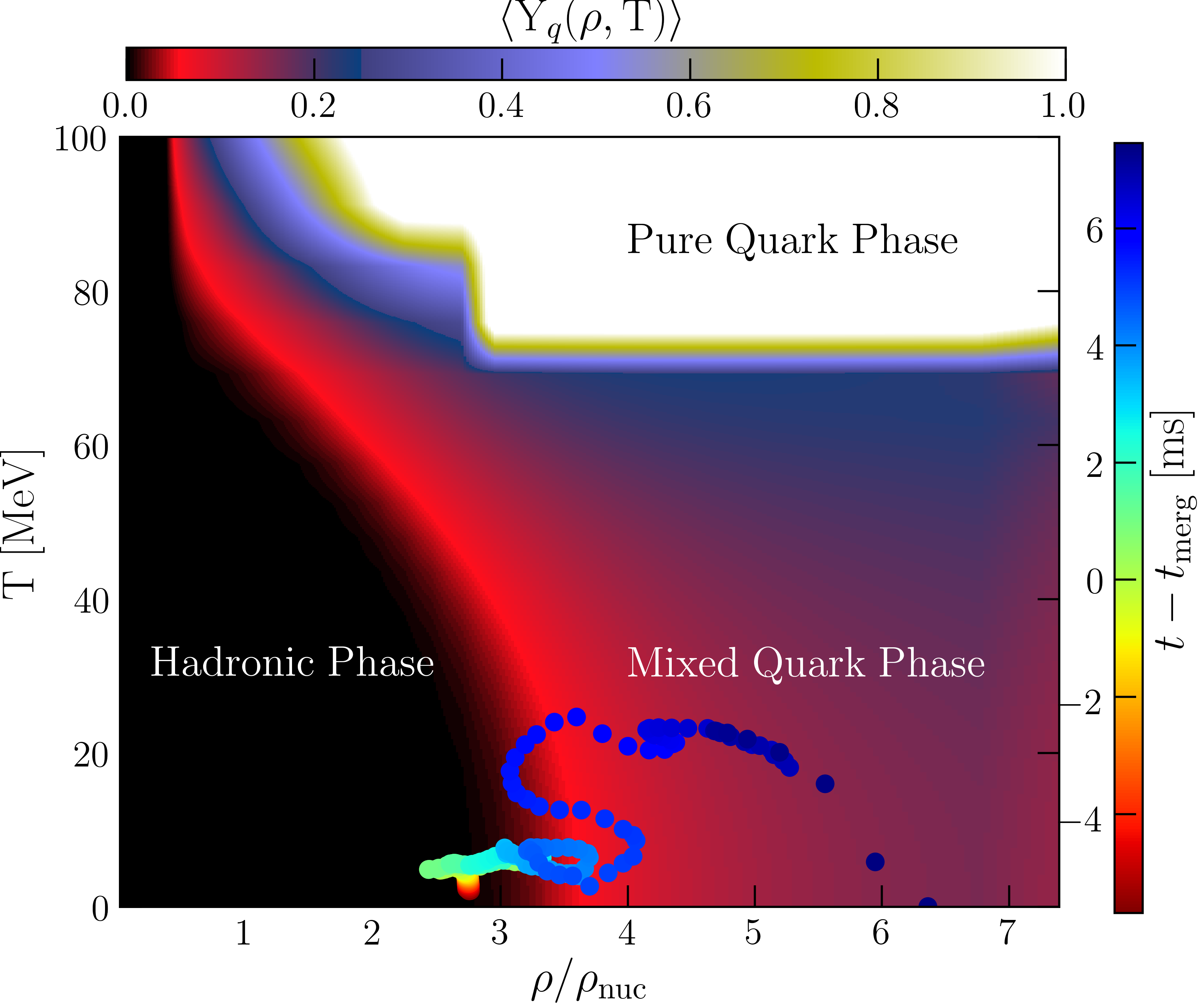}
  \includegraphics[width=0.49\textwidth]{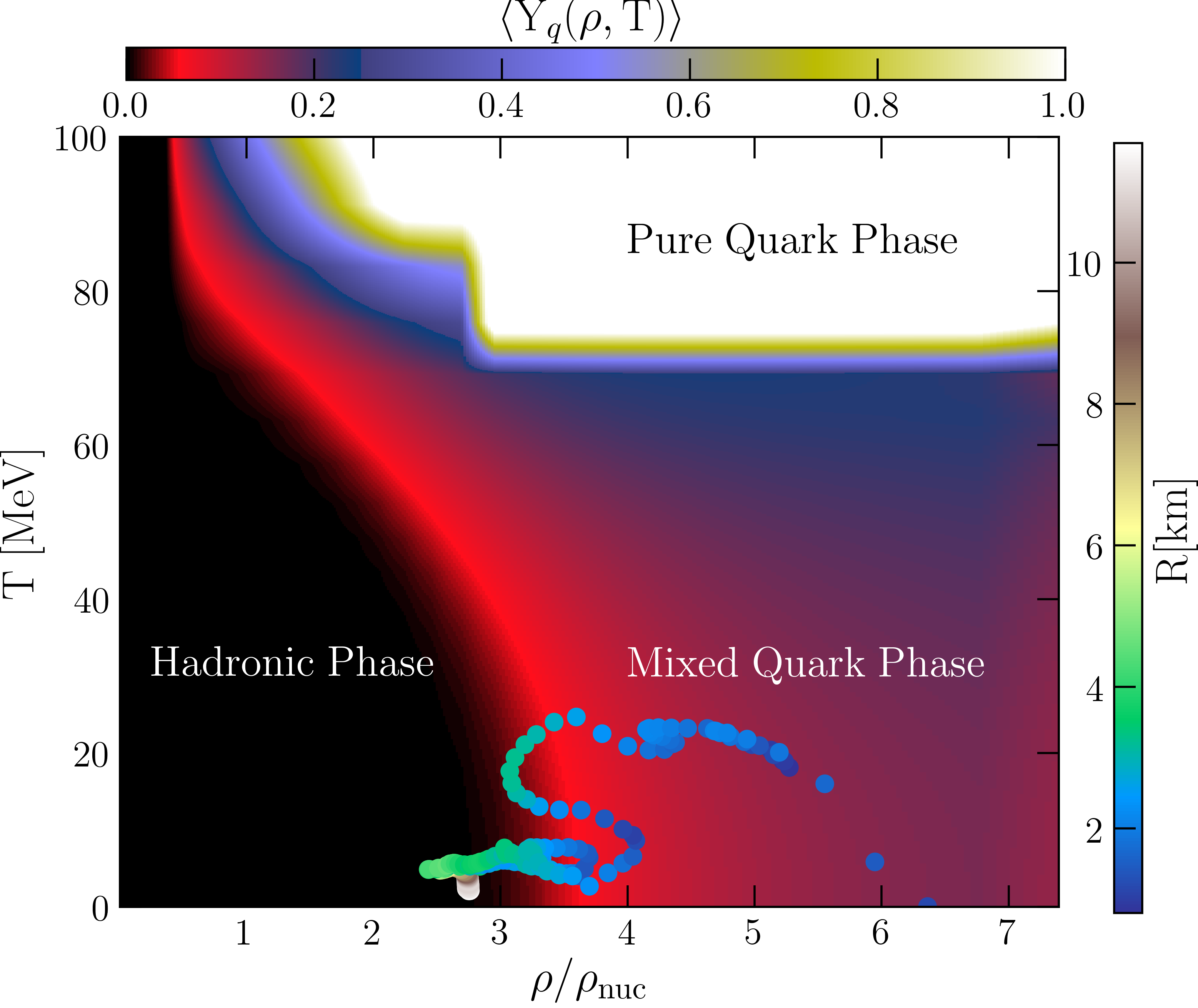}
  \caption{Thermodynamic trajectory of a representative tracer particle from the binary system $1.3325\ \msun - 1.3325\ \msun$. The trajectory is superposed on a $\mathrm{Y_e}$ weighted equilibrium slice of the BLQ EOS. The trajectories themselves are color coded according to the relative time from merger and the radial distance of a tracer from the center of the remnant. Matter in the NS cores crosses the phase boundary several times starting from the moment of merger and until the time of collapse and BH formation.}
  \label{fig:probe}
\end{figure*}

The result of this procedure constitutes the background color map used in figure \ref{fig:probe}. The figure also shows the thermodynamic trajectory of a Lagrangian tracer particle from our fiducial binary. This particle is representative of the evolution of the tracer particles that are located in the inner 7.3 km of the remnant. Before the merger, corresponding to negative times and coordinate distances larger than $\sim 7\ {\rm km}$, the particle is in the hadronic phase. As the stars merge, matter undergoes repeated cycles of compression and expansion, during which it crosses the phase boundary repeatedly. After a few oscillations, the remnant becomes unstable and starts to collapse. Matter is compressed to large densities and BH formation takes place. As the tracer evolves, the radial distance oscillates between $4~\rm{km}$ and $6~\rm{km}$ from the remnant's centre, signifying that as the tracer is moving in and out of the quark phase, it is also moving in and around the remnant's core. Finally, just before the final collapse, the tracer particle is found in the hadron-quark  mixed phase of the EOS. The occurrence of mixed quark phases is a consequence of modelling the phase transition by imposing global charge neutrality or the Gibb's construction. Gibb's construction was also utilized in the work by Blacker et al. \cite{Blacker:2020nlq} which resulted in mixed quark phases with different onset densities (at $T=0$) in their quark EOSs. This is an important difference with respect to the models used in some of the earlier works \cite{Most:2018eaw, Most:2019onn}, in which the transition to pure quark matter occurred within a narrow range of densities and temperatures.

\begin{figure}
\includegraphics[width=\columnwidth]{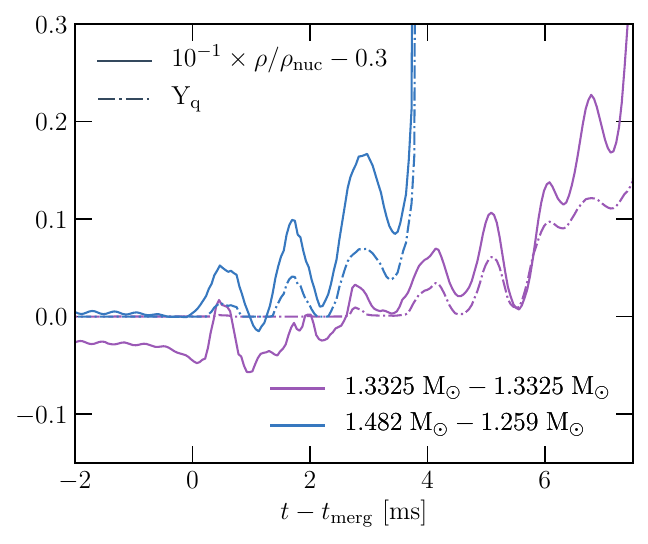}
\caption{\label{fig:Yq_rho_t} Time evolution of quark fraction and density of fluid elements traced by Lagrangian tracers for 2 binary neutron star systems 1.482 $\msun$-1.259 $\msun$ and 1.3325-1.3325 $\msun$. Noticeable is the fact that the period of oscillations of density matches the period of oscillations of quark fraction.}
\end{figure}

The presence of matter repeatedly crossing the phase boundary is generic across our set of simulations. Such a behavior was expected to take place on the basis of the analysis of thermodynamic trajectories from simulations that employed purely hadronic EOSs \cite{Hanauske:2019qgs}. It is now confirmed for the first time in our work. This process is also illustrated in figure \ref{fig:Yq_rho_t}, where we show the evolution in time for the quark fraction and density for two tracer particles: one from the $1.3325\ \msun - 1.3325\ \msun$ binary and one from the $1.482\ \msun - 1.259\ \msun$ binary. We find that the oscillations in $\rm{Y_q}$ correlate with the density oscillations. This suggests that the phase boundary traversal is triggered by the oscillations of the remnant after its formation. Indeed, we find that the density oscillations in the tracer particles closely track those of the maximum density. These density oscillations either culminate in BH formation or are damped over a timescale of ${\sim}10{-}20\ {\rm ms}$ \cite{Bernuzzi:2015opx}.

\begin{figure}
\includegraphics[width=\columnwidth]{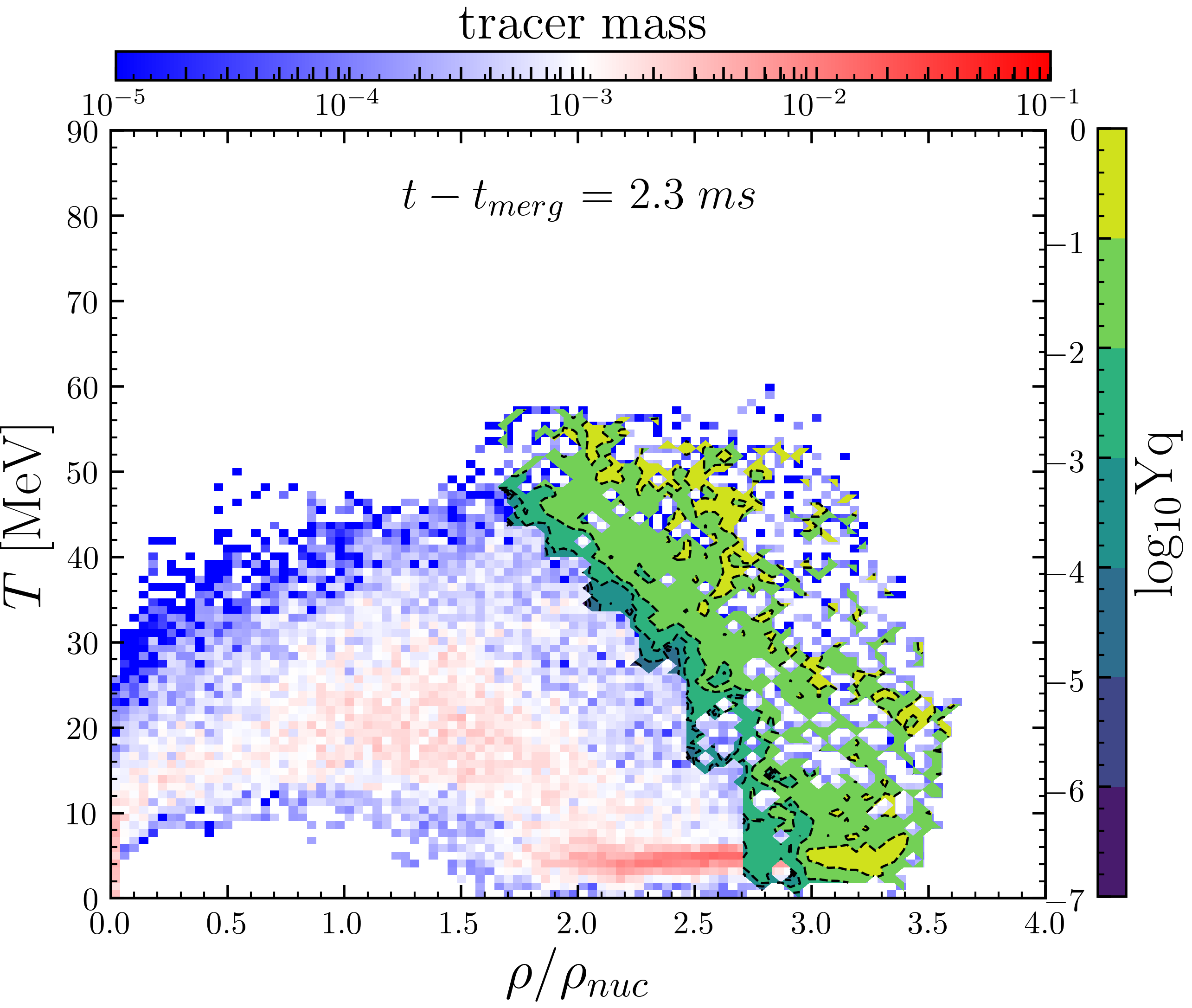}
\caption{\label{fig:2D_Hist} A two-dimensional histogram of the thermodynamic variables $\mathrm{\rho}$ and $\mathrm{T}$ and weighted by bins of tracer mass. Also shown are contours of quark fraction. Both the bulk of the remnant's core and the periphery of the core can exhibit deconfined quark matter depending upon $\mathrm{\rho}$ and $\mathrm{T}$.}
\end{figure}

Figure \ref{fig:2D_Hist} shows a complementary analysis of the phase diagram for the $1.3325\ \msun - 1.3325\ \msun$ binary. Instead of showing the thermodynamic trajectories of a specific fluid element, we provide a snapshot of the entire star at a fixed time, 2.3~ms after the merger. In particular, we show a histogram of the data from all the tracers in the simulation. The color represents the sum of all the masses of tracer particles in a particular temperature and density range. We also plot contours of the quark fraction on and above the threshold for the deconfinement phase transition. The red streak at low temperature ${\sim}10\ {\rm MeV}$ in the range of densities between $2\rho_{\rm nuc}{-}3 \rho_{\rm nuc}$ represents the thermodynamic state of the bulk of the remnant's core matter. This extends to sufficiently high densities to cross the phase boundary according to our adopted EOS. However, quarks are also formed at lower densities in the high temperature regions ($\rm{T} \sim 50\ {\rm MeV}$) that have been heated during the merger phase. These hot layers are located at the periphery of the quark core \cite{Bernuzzi:2015opx}. Our data show that, if indeed the phase transition to deconfined quarks is of the first order, then, depending on the particular location of the critical point in the QCD phase diagram, matter could simultaneously undergo both a first order and a crossover phase transition in different parts of the remnant. We remark here that these hot and cold regions of deconfined quark matter in and around the remnant's core were also observed by Blacker et al. \cite{Blacker:2020nlq} in their rest mass distributions over a density-temperature plane of DD2F-SF EOSs.

We define certain characteristic times in our simulations as follows. $t_{\rm{merg}}$ is the time of merger, taken to be the time when the amplitude of the $l=2, m=2$ mode of the GW strain attains a maximum. This point approximately coincides with a minima in the maximum density of the stars, as the NSs are plunging towards each other \cite{Baiotti:2008ra}. This expansion phase is followed by a very rapid and strong compression as the stars collide (see figure~\ref{fig:h22}). $t_{\rm{BH}}$ is the time of formation of a black hole in the simulation marked by the formation of an apparent horizon, which we approximately take to be the time when the lapse function drops below 0.2.  We denote by $t_{\rm{coll}}$ the time when the $l=2, m=2$ mode of the radiation effectively shuts off which we take to be the time when the amplitude drops below 0.5 $\%$ of its maximum value. $t_{\rm{end}}$ is the terminal point of our simulations where we cease to evolve the system. These times scales are reported in milliseconds from merger in table \ref{tab:GW_property}.


\section{Gravitational Waves}
\label{sec:gravitatoinal_waves}

\begin{figure*}
  \centering
  \includegraphics[width=0.49\textwidth]{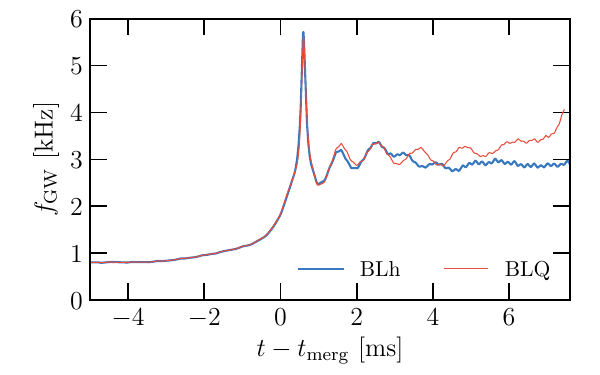}
  \includegraphics[width=0.49\textwidth]{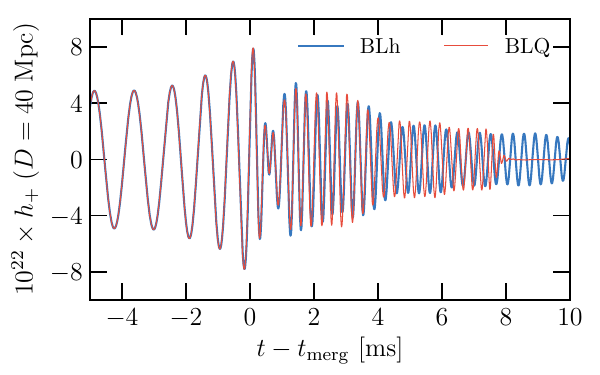}
  \includegraphics[width=0.49\textwidth]{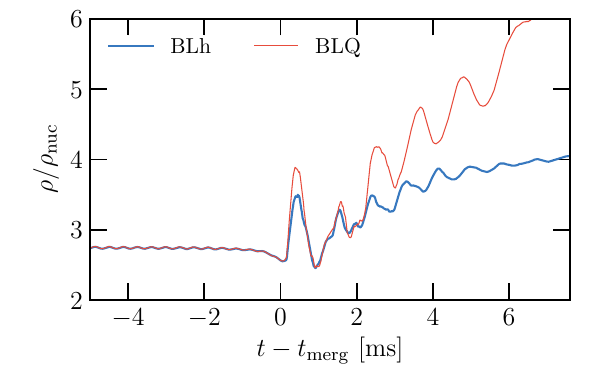}
  \includegraphics[width=0.49\textwidth]{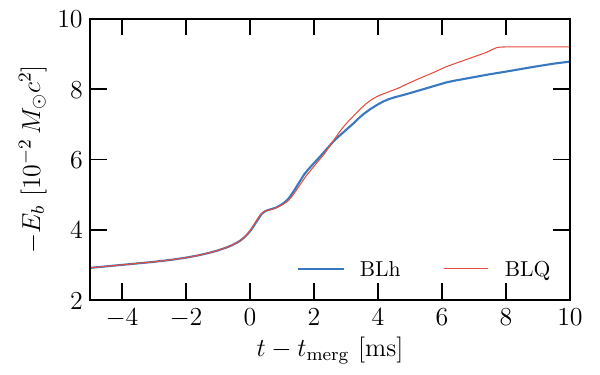}
  \caption{Evolution of the instantaneous GW frequency $f_{\rm{GW}}$, the``+'' polarization strain amplitude for the ($l=2, m=2$) mode of the GW signal, the central density $\rho$, and the binding energy $E_b$ of the $1.3325\ \msun - 1.3325\ \msun$ binary. The inspiral ($t \leq t_{\rm{merg}}$) evolution predicted by both the BLh and BLQ EOSs is identical. The appearance of quarks is imprinted on the postmerger dynamics and GW signal.}
  \label{fig:h22}
\end{figure*}

Figure~\ref{fig:h22} shows the strain amplitude and instantaneous frequency of the dominant $(l=2, m=2)$ mode of the GW strain for the $1.3325\ \msun - 1.3325\ \msun$ binary, as well as the maximum density, and the binary binding energy. The latter is computed by subtracting the energy radiated in GWs from the initial binding energy of the binary $\rm{M}_{\rm ADM} - (\rm{M}_{1} + \rm{M}_{2})$ following \cite{Bernuzzi:2012ci, Bernuzzi:2015opx}. We find that the inspiral dynamics for the BLQ binary is identical to that of the BLh binary. This is expected, because for this binary, as well as for most of the binaries considered in this study, the two EOSs are identical over the range of densities and temperatures reached in the inspiral. Nevertheless, this is an important consistency check, given that we simulate both the BLh and BLQ binaries starting from the same initial data. That is, we do not start the BLQ simulations from pre-merger snapshots of the BLh simulations.

\begin{figure*}
    \centering
    \includegraphics[width=0.32\textwidth]{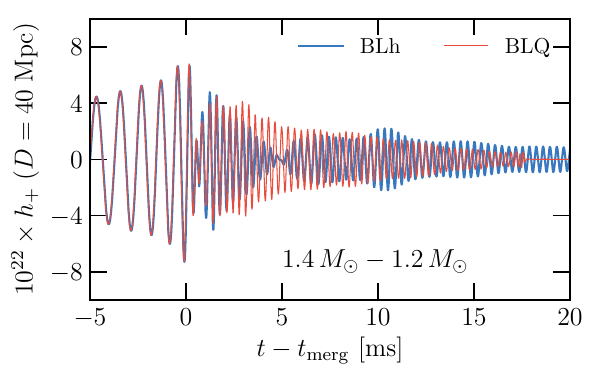}
    \includegraphics[width=0.32\textwidth]{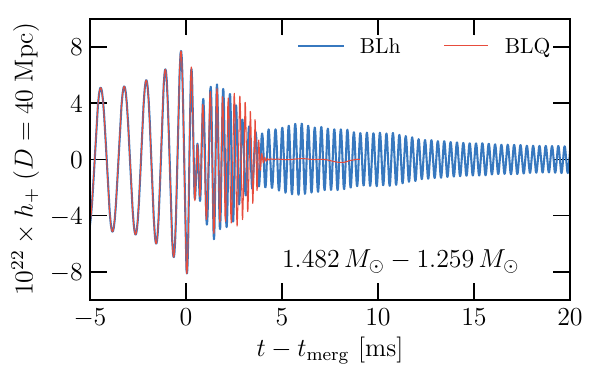}
    \includegraphics[width=0.32\textwidth]{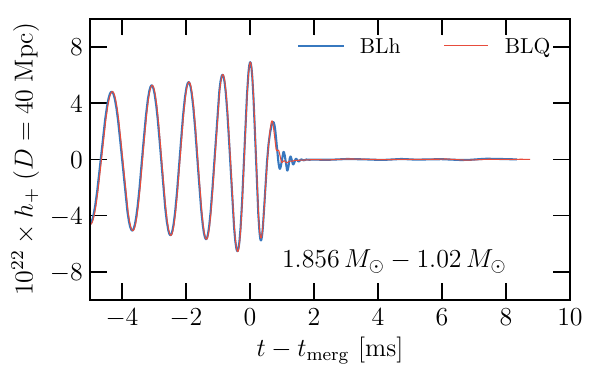}
    \includegraphics[width=0.32\textwidth]{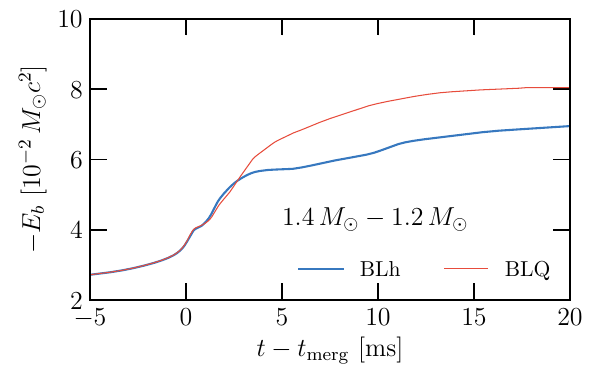}
    \includegraphics[width=0.32\textwidth]{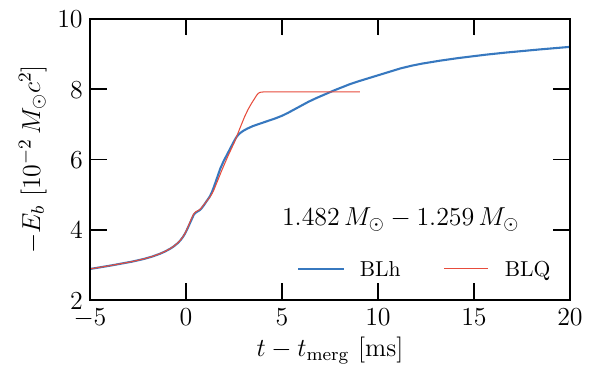}
    \includegraphics[width=0.32\textwidth]{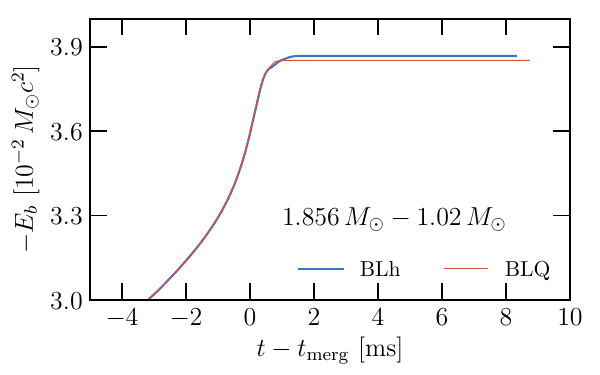}
    \caption{Amplitude of the ($l=2, m=2$) mode of the GW strain $h_{+}$ and binding energies for the $1.4\ \msun$ - $1.2\ \msun$, $1.482\ \msun$ - $1.259\ \msun$, and $1.856\ \msun$ - $1.020\ \msun$ binaries. As the binaries become more massive or more asymmetric, the length of the postmerger signal decreases. The postmerger is further shortened by an onset of deconfinement phase transition.}
    \label{fig:h22_2}
\end{figure*}

The only exception is the $1.856\ \msun - 1.020\ \msun$ binary (figure \ref{fig:h22_2}), in which quarks are already present during the inspiral according to the BLQ EOS. For this binary we find the pre-merger maximum quark fraction to be $\rm{Y_q} \simeq 0.06 $. The maximum density in the BLQ binary is $\sim 7\%$ larger than that of the BLh binary. Despite these differences, the orbital evolution for the $1.856\ \msun - 1.020\ \msun$ BLh and BLQ binaries are essentially indistinguishable. In particular, the dephasing between the two associated waveforms is smaller than our numerical precision. This is also not surprising given that the phase transition only impacts the primary component of this binary, whose tidal parameter $\Lambda_1$ is in any case very small $\simeq 45$.

The postmerger evolution of the BLh and BLQ binaries are instead very different. This is shown in figure~\ref{fig:h22} for the $1.3325\ \msun - 1.3325\ \msun$ binaries. The phase transition to deconfined quarks in the BLQ binaries is accompanied by a loss of pressure support. This in turn causes the BLQ remnants to undergo violent cycles of gravitational contraction and centrifugal bounces, while at the same time becoming progressively more compact. This ultimately leads to the collapse to BH. The progressive contraction of the remnant is accompanied by a drift in the instantaneous peak frequency of the $(l=2, m=2)$ mode of the GW signal and by an increase in the overall GW luminosity, as evidenced by the evolution of $E_b$ in figure~\ref{fig:h22}. 

Figure~\ref{fig:h22_2} shows the general trends of the $(l=2, m=2)$ component of the GW strain and of the binary binding energies across our set of simulations. We do not find any significant difference in the inspiral GW signal between the BLh and BLQ binaries. This also includes the $1.856\ \msun$ - $1.020\ \msun$ binary for which quarks are also present in the inspiral, as discussed above. Significant differences are present in the postmerger for all the binaries, with the exception of the $1.856\ \msun$ - $1.020\ \msun$ binary, which results in prompt BH formation. For the latter, the postmerger signal is consistent with the ring down of the formed BH. The phase transition is imprinted in the duration of the postmerger signal and as a change in the peak frequency and overall amplitude of the signal. All BLQ binaries form BHs during our simulation time (see also table \ref{tab:GW_property}). The change in the amplitude of the GW signal is similar to that reported by Radice et al.~\cite{Radice:2016rys}, who studied the impact of the appearance of hyperons in mergers. However, in the case of a first order transition, the change in the amplitude is accompanied by a change in the peak frequency of the postmerger signal \cite{Bauswein:2018bma}.

More specifically, the works by Sekuguchi et al.\cite{Sekiguchi:2011mc} and Radice et al. \cite{Radice:2016rys} found that the appearance of hyperons lead to a softening of the EOS which is qualitativaly similar to that induced by a first order quark deconfinement phase transition and documented here. Indeed, like the quark deconfinement phase transition, the creation of hyperons lead to more compact remnants that are more prone to collapse. However, because the thermodynamical potentials of matter remain smooth, the appearance of hyperons does not impact the peak frequency of the postmerger signal of the remnants, if not for the fact that binaries simulated with hyperons typically results in earlier BH formation.

\begin{figure*}
  \centering
  \includegraphics[width=0.49\textwidth]{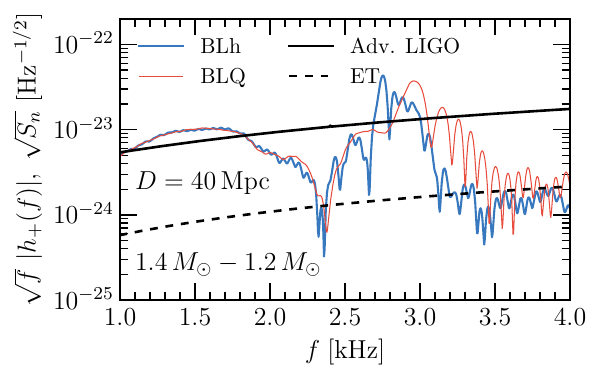}
  \includegraphics[width=0.49\textwidth]{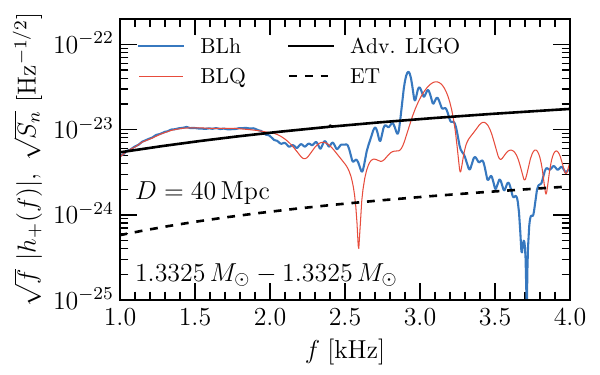}
  \includegraphics[width=0.49\textwidth]{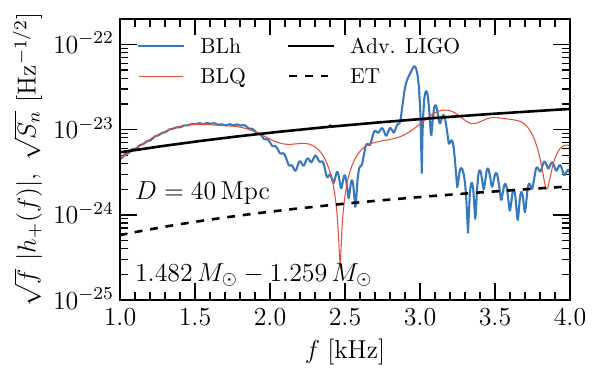}
  \includegraphics[width=0.49\textwidth]{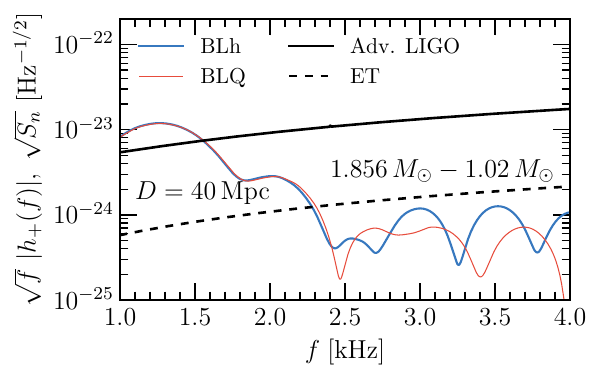}
  \caption{Power spectrum of the $(l=2, m=2)$ mode of the GW strain for the $1.4\ \msun - 1.2\ \msun$, $1.3325\ \msun - 1.3325\ \msun$, $1.482\ \msun - 1.259\ \msun$, and $1.856\ \msun - 1.020\ \msun$ binaries. An exponential filter was applied to the data to remove the inspiral signal. The difference in the peak frequency between the BLQ and the BLh binaries in the top panels is sufficiently large to be measured. On the other hand, because of the short length of the BLQ postmerger signals, the differences in the peak frequency for the binaries in the bottom panels is smaller than the nominal uncertainty of the Fourier transform, so they cannot be measured.}
  \label{fig:PSD}
\end{figure*}

The GW power spectra for a representative set of binaries are shown in figure~\ref{fig:PSD}. A characteristic feature in the postmerger spectra of NS mergers is the existence of a postmerger peak frequency $f_2$ \cite{Takami:2014tva, Bernuzzi:2015rla, Dietrich:2016hky, Dietrich:2016lyp, Shibata:2002jb, Stergioulas:2011gd, Bauswein:2011tp, Hotokezaka:2013iia, Takami:2014zpa, Radice:2016gym, Lehner:2016lxy} in the range of 2-4~kHz that is related with the rotational frequency of the remnant. As is evident from figure~\ref{fig:PSD}, there is a characteristic shift in this postmerger peak frequency due to the appearance of quarks in binaries evolved with the BLQ EOS. This is in qualitative agreement with the findings of Bauswein et al. \cite{Bauswein:2018bma}, who found that such shift is a distinctive signature of a phase transition. However, the maximum shift in $f_2$ among our simulations is only 0.21~kHz (table \ref{tab:GW_property}). This is to be contrasted with the range of shifts (0.2-0.6 kHz) in $f_2$ observed by Bauswein et al. \cite{Bauswein:2018bma}. These difference could arise due to the differences in the EOS models used in this study and in Bauswein et al. ~\cite{Bauswein:2018bma}, or they could be due to differences in the methodologies of our simulations. We cannot verify this since the EOS models used by Bauswein et al.~\cite{Bauswein:2018bma} are not available to us.

We remark that these shifts in the postmerger peak frequencies can only be observed for binaries with a sufficiently long postmerger, where the Fourier uncertainty principle would imply a finite localization of power spectral density (PSD) peaks in the frequency domain. For a short-lived remnant, the uncertainty principle dictates that there would be a spread of PSD over a wide range of frequencies and hence the peaks would be too broad to observe any shifts (See figure \ref{fig:PSD}). The uncertainty in the Fourier transform is the reciprocal of the length of the postmerger signal, i.e., $\Delta \rm{FT}$ $= 1/(t_{\rm{coll}} - t_{\rm{merg}})$. For long lived remnants, i.e., $1.30\ \msun - 1.30\ \msun$, $1.3325\ \msun - 1.3325\ \msun$ and $1.4\ \msun - 1.2\ \msun$, the shift in peak frequency between BLh and BLQ EOS satisfies $\Delta f_2 > \Delta \rm{FT}$ and hence $f_2$ shift can be a robust signature of a phase transition. For shorter lived remnants, like $1.482\ \msun - 1.259\ \msun$ and $1.365\ \msun - 1.365\ \msun$, $\Delta f_2 \leq \Delta \rm{FT}$ and so the frequency shifts are not good indicators of phase transitions here. Finally, for systems with very little to no postmerger, like the equal mass binaries from $1.4\ \msun - 1.4\ \msun$ to $1.6\ \msun - 1.6\ \msun$, and $1.856\ \msun - 1.020\ \msun$, $t_{\rm{coll}} \to t_{\rm{merg}}$, so the frequency distribution of the postmerger signal is too broad to compute any robust signature. We find that this criterion for comparing the shifts in $f_2$ with uncertainties in the Fourier transform of the time domain signal holds across the two spatial resolutions we have investigated (see table \ref{tab:GW_property}).

\begin{figure}
  \centering
  \includegraphics[width=0.98\columnwidth]{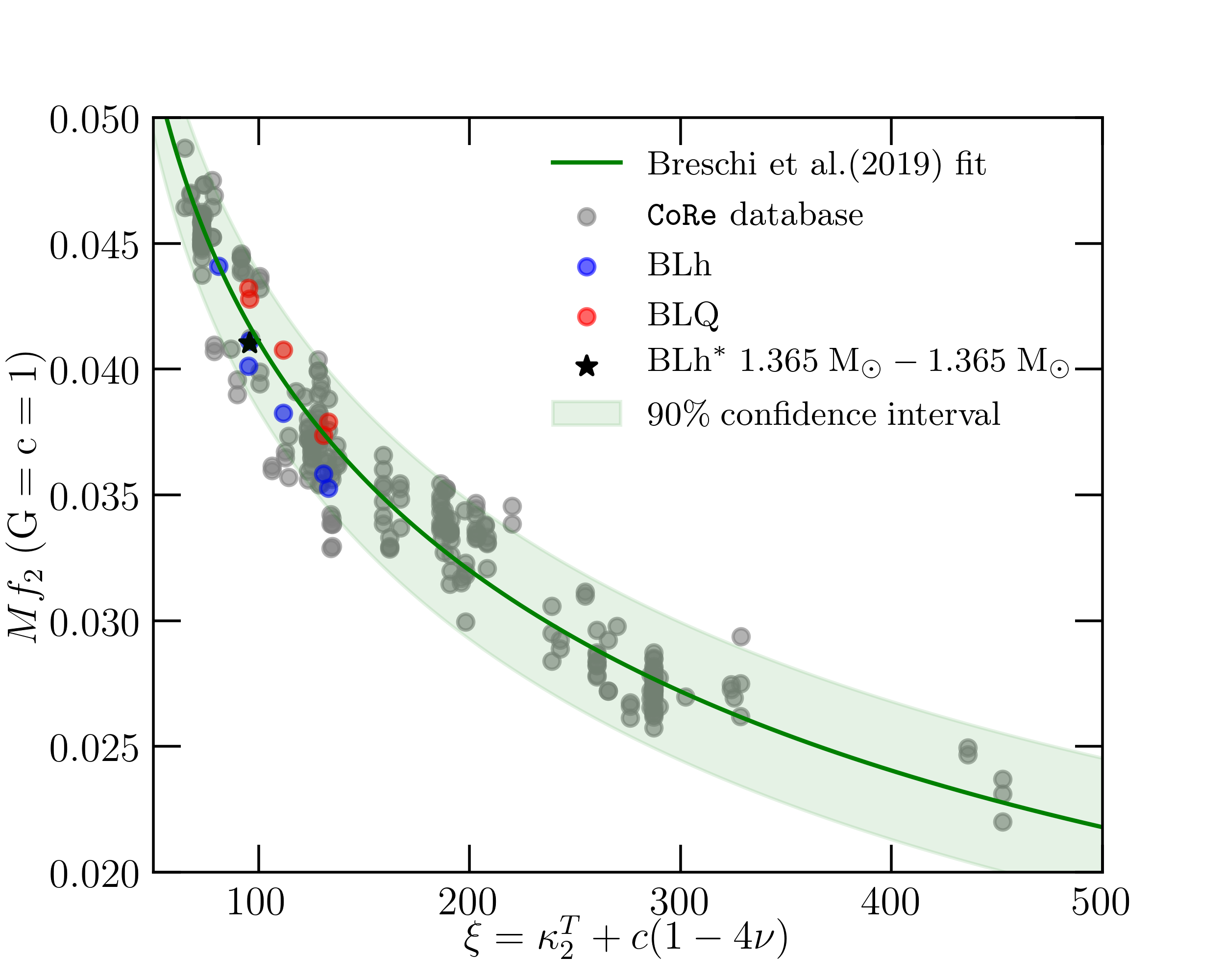}
  \caption{Correlations between the total mass-scaled postmerger peak frequency $Mf_2$ and the tidal parameter $\xi$. Also shown is the fit from the quasi universal relation presented in \cite{Breschi:2019srl} along with its $90\%$ confidence interval. The grey points correspond to simulations catalogued in the $\tt{CoRe}$ database \cite{Dietrich:2018phi}. It can be seen that deviations in $f_2$ (red circles) by virtue of phase transitions are not large enough to violate the quasi-universal relation.}
  \label{fig:fit}
\end{figure}

To eliminate additional sources of these $f_2$ shifts, we perform a simulation of the $1.365\ \msun - 1.365\ \msun$ binary (table \ref{tab:GW_property}) using the subgrid-scale turbulence model of Radice~\cite{Radice:2020ids}, which was calibrated using the GRMHD simulations of Kiuchi et al.~\cite{Kiuchi:2017zzg}. We find that the introduction of viscosity can result in the appearance of secondary peaks in the postmerger spectrum that are formed in the first few milliseconds after the postmerger. However, the $f_2$ peak frequency is not affected, in agreement with our previous findings \cite{Radice:2017zta}. In this paper, we denote the results from this run with an asterisk to the EOS name as BLh* (see tables \ref{tab:GW_property}, \ref{tab:ejecta}, and figures \ref{fig:fit}, \ref{fig:disk_mass}).

The $f_2$ peak frequency has been shown to be tightly correlated with the NS radius and tidal properties of a binary \cite{Bauswein:2012ya, Hotokezaka:2013iia, Bernuzzi:2014kca, Rezzolla:2016nxn, Zappa:2017xba, Bauswein:2011tp, Lioutas:2021jbl, Bernuzzi:2015rla}. These empirical,  quasi-universal relations are interesting because they correlate the tidal properties of a binary, which are characteristic of the inspiral, to the postmerger peak frequency $f_2$ of the remnant. A significant departure from these relations caused by shifts in $f_2$ can provide conclusive evidence for deconfinement phase transitions \cite{Bauswein:2018bma, Bauswein:2020ggy, Weih:2019xvw, Blacker:2020nlq}. We test the quasi-universal relation proposed in Breschi et al. \cite{Breschi:2019srl} against our $f_2$ frequencies and present the results in figure \ref{fig:fit}. In line with the terminology presented in \cite{Breschi:2019srl}, we plot our postmerger peak frequencies from BLh and BLQ binaries against the tidal parameter $\xi$ defined by:
\begin{equation}
    \xi = \kappa_{2}^{T} + c(1 - 4 \nu) \, ,
\end{equation}
where $c$ is a fitting parameter, $\nu = M_1 M_2 / (M)^2$ is the symmetric mass ratio, and $\kappa_{2}^{T}$ is the tidal polarizability parameter defined by
\begin{equation}
    \kappa_{2}^{T} = 3 \left(\Lambda_1 X_1^4 X_2 + \Lambda_2 X_2^4 X_1 \right).
\end{equation}
Here $X_i = M_i/M$. The functional form of the quasi-universal relation is given as:
\begin{equation}
    Mf_2 = F_0\frac{1 + n_1\xi + n_2\xi^2}{1 + d_1\xi + d_2\xi^2}
    \label{eq:fitt}
\end{equation}
where $F_0, n_1, n_2, d_1, d_2$ and $c$ are fitting parameters specified in \cite{Breschi:2019srl}.
As mentioned previously, the shifts in $f_2$ (when observed i.e. when $\Delta f_2 > \Delta \rm{FT}$) have been very modest as compared to Bauswein et al. \cite{Bauswein:2018bma, Bauswein:2020ggy} and Blacker et al. \cite{Blacker:2020nlq} and are found to be described very well by Eq.~(\ref{eq:fitt}). We do not find any evidence of strong deviations from the fit as a consequence of deconfinement phase transitions. Our results show that the absence of deviations in the $f_2$ from the expected universal relations cannot be used to rule out the presence of phase transitions, or to constrain their density threshold, as instead claimed by Blacker et al. \cite{Blacker:2020nlq}. Additionally, we would like to remark that the quasi-universal relation presented in figure \ref{fig:fit} describes a larger dataset of BNS mergers than any of the previous studies by including simulations from 14 EOSs (including hyperonic and quark EOSs) and a large sampling of mass ratios ranging from 1 to 2.06 to describe asymmetric binaries which are consistently taken into account by the parameter $\xi$. Finally, we caution the reader that, even though shifts in $f_2$ have not been reported in the literature for purely hadronic EOSs, we cannot exclude the possibility that such shifts might occur for hadronic EOSs exhibitting a sudden change in their stiffness at several times the saturation density.



\section{Dynamical Ejecta and Disks}
\label{sec:dynamical_ejecta_and_disks}
\subsection{Ejecta and Nucleosynthesis}
\label{subsec:ejecta_annd_nucleosynthesis}

We now describe the properties of the outflow from a merger with a deconfinement phase transition which will eventually help calculate possible EM counterparts of such mergers. To this aim, we calculate the asymptotic velocity, the specific entropy, the angle with the orbital plane, and the electron fraction of the matter ejected on a dynamical timescale in our simulations. In particular, we record the properties of matter that is crossing a coordinate sphere of radius 200 $G \msun / c^2$ ($\simeq 295.34$~km) and that is unbound according to the geodesic criterion, i.e., with $u_t<-1$. We refer to \cite{Kastaun:2014fna, Bovard:2017mvn, Nedora:2019jhl} for a discussion of other possible choices.

\begin{table*}[t] 
\begin{center} 
\caption{ 
 A summary of the analysis of ejecta properties and disk masses for all our binaries at 2 spatial resolutions. $\rm{M}_{\text{disk}}^{\text{end}}$ is the disk mass at the end of the simulation, $\rm{M}_{\text{ej}}$ is the total mass of the ejecta, $\langle v_{\infty}\rangle_{ej}$ is the ejecta's mass averaged asymptotic velocity, $\langle \rm{Y_e} \rangle_{ej}$ its mass averaged electron fraction, $\langle s \rangle_{ej}$ the mass averaged specific entropy and $\langle \theta \rangle_{ej}$ is the rms angle with the orbital plane. 
} 
\label{tab:ejecta} 
\scalebox{1.05}{ 
 \begin{tabular}{c c c c c c c c c c c c c c} 
 \hline\hline 
 \\[-0.9em] 
 EOS & $\rm{M_1}$ & $\rm{M_2}$ & M & q & Resolution & $\rm{M}_{\text{disk}}^{\text{end}}$ & $\rm{M}_{\text{ej}}$ & $\langle v_{\infty}\rangle_{ej}$ & $E_{\text{kin}}$ & $E_{\text{kin}}(W\beta>1)$ & $\langle \rm{Y_e} \rangle_{ej}$ & $\langle s \rangle_{ej}$ & $\langle \theta \rangle_{ej}$ \\ 
 & [$\msun$] & [$\msun$] & [$\msun$] & & & $[10^{-3} \msun]$ & $[10^{-3} \msun]$ & [c] & $[10^{48} \rm{erg}]$ & $[10^{48} \rm{erg}]$ &  & [$k_B$] & [$\rm{rad}$] \\ 
 \hline 
 \hline 
BLh & 1.3 & 1.3 & 2.6 & 1.0 & SR & 110.29 & 1.22 & 0.14 & 34.48 & 0.22 & 0.25 & 21.51 & 0.65\\
BLQ & 1.3 & 1.3 & 2.6 & 1.0 & SR & 58.89 & 1.73 & 0.15 & 50.55 & 0.23 & 0.25 & 21.67 & 0.61\\
\hline 
BLh & 1.3325 & 1.3325 & 2.67 & 1.0 & SR & 83.44 & 0.88 & 0.18 & 37.98 & 1.31 & 0.22 & 20.13 & 0.61\\
BLQ & 1.3325 & 1.3325 & 2.67 & 1.0 & SR & 16.79 & 1.18 & 0.21 & 69.4 & 4.62 & 0.21 & 18.61 & 0.57\\
\hline 
BLh & 1.365 & 1.365 & 2.73 & 1.0 & SR & 49.44 & 1.4 & 0.16 & 45.3 & 0.16 & 0.26 & 21.14 & 0.65\\
BLh* & 1.365 & 1.365 & 2.73 & 1.0 & SR & 53.35 & 1.51 & 0.2 & 75.47 & 3.33 & 0.25 & 22.16 & 0.64\\
BLQ & 1.365 & 1.365 & 2.73 & 1.0 & SR & 6.95 & 2.05 & 0.2 & 97.53 & 0.76 & 0.24 & 18.56 & 0.6\\
\hline 
BLh & 1.4 & 1.4 & 2.8 & 1.0 & SR & 80.67 & 1.85 & 0.17 & 65.82 & 0.92 & 0.25 & 21.1 & 0.62\\
BLQ & 1.4 & 1.4 & 2.8 & 1.0 & SR & 2.99 & 0.49 & 0.23 & 31.3 & 1.79 & 0.21 & 20.52 & 0.47\\
\hline 
BLh & 1.45 & 1.45 & 2.9 & 1.0 & SR & 76.05 & 6.84 & 0.19 & 309.73 & 7.7 & 0.25 & 17.03 & 0.57\\
BLQ & 1.45 & 1.45 & 2.9 & 1.0 & SR & 0.13 & 0.12 & 0.26 & 8.36 & 0.0 & 0.25 & 28.65 & 0.4\\
\hline 
BLh & 1.475 & 1.475 & 2.95 & 1.0 & SR & 0.04 & 0.33 & 0.29 & 31.56 & 0.87 & 0.22 & 22.68 & 0.41\\
BLQ & 1.475 & 1.475 & 2.95 & 1.0 & SR & 0.13 & 0.04 & 0.25 & 2.69 & 0.0 & 0.26 & 43.51 & 0.36\\
\hline 
BLh & 1.5 & 1.5 & 3.0 & 1.0 & SR & 0.1 & 0.2 & 0.29 & 19.21 & 0.3 & 0.24 & 23.58 & 0.32\\
BLQ & 1.5 & 1.5 & 3.0 & 1.0 & SR & 0.04 & 0.01 & 0.15 & 0.19 & 0.0 & 0.19 & 65.38 & 0.34\\
\hline 
BLh & 1.6 & 1.6 & 3.2 & 1.0 & SR & 0.01 & 0.0 & 0.16 & 0.1 & 0.0 & 0.22 & 91.26 & 0.35\\
BLQ & 1.6 & 1.6 & 3.2 & 1.0 & SR & 0.01 & 0.0 & 0.17 & 0.13 & 0.0 & 0.22 & 87.72 & 0.36\\
\hline 
BLh & 1.4 & 1.2 & 2.6 & 1.17 & SR & 107.27 & 1.98 & 0.19 & 92.57 & 2.39 & 0.18 & 14.06 & 0.55\\
BLQ & 1.4 & 1.2 & 2.6 & 1.17 & SR & 67.01 & 1.82 & 0.2 & 96.53 & 4.09 & 0.17 & 14.09 & 0.5\\
\hline 
BLh & 1.482 & 1.259 & 2.74 & 1.18 & SR & 92.76 & 4.96 & 0.17 & 178.19 & 1.5 & 0.17 & 12.33 & 0.5\\
BLQ & 1.482 & 1.259 & 2.74 & 1.18 & SR & 13.44 & 2.53 & 0.22 & 154.29 & 2.73 & 0.14 & 11.53 & 0.45\\
\hline 
BLh & 1.856 & 1.02 & 2.88 & 1.82 & SR & 60.99 & 7.51 & 0.11 & 102.35 & 0.15 & 0.04 & 3.8 & 0.11\\
BLQ & 1.856 & 1.02 & 2.88 & 1.82 & SR & 59.46 & 7.42 & 0.11 & 101.71 & 0.08 & 0.04 & 4.04 & 0.12\\
\hline 
\hline 
BLh & 1.3 & 1.3 & 2.6 & 1.0 & LR & 87.61 & 0.95 & 0.15 & 27.89 & 0.0 & 0.24 & 22.4 & 0.62\\
BLQ & 1.3 & 1.3 & 2.6 & 1.0 & LR & 28.71 & 0.93 & 0.19 & 42.01 & 0.28 & 0.23 & 21.41 & 0.63\\
\hline 
BLh & 1.3325 & 1.3325 & 2.67 & 1.0 & LR & 95.56 & 1.28 & 0.18 & 50.3 & 0.07 & 0.26 & 22.91 & 0.64\\
BLQ & 1.3325 & 1.3325 & 2.67 & 1.0 & LR & 43.02 & 1.16 & 0.17 & 40.1 & 0.0 & 0.27 & 23.53 & 0.64\\
\hline 
BLh & 1.365 & 1.365 & 2.73 & 1.0 & LR & 65.21 & 1.11 & 0.16 & 36.01 & 0.29 & 0.27 & 25.36 & 0.67\\
BLQ & 1.365 & 1.365 & 2.73 & 1.0 & LR & 6.35 & 2.29 & 0.17 & 84.07 & 0.01 & 0.26 & 20.48 & 0.59\\
\hline 
BLh & 1.4 & 1.4 & 2.8 & 1.0 & LR & 72.59 & 1.68 & 0.19 & 77.06 & 0.41 & 0.26 & 20.11 & 0.67\\
BLQ & 1.4 & 1.4 & 2.8 & 1.0 & LR & 9.07 & 0.51 & 0.25 & 37.22 & 0.43 & 0.23 & 22.66 & 0.53\\
\hline 
BLh & 1.45 & 1.45 & 2.9 & 1.0 & LR & 2.42 & 1.5 & 0.23 & 95.62 & 0.91 & 0.26 & 20.46 & 0.52\\
BLQ & 1.45 & 1.45 & 2.9 & 1.0 & LR & 0.06 & 0.09 & 0.26 & 6.96 & 0.0 & 0.26 & 33.35 & 0.39\\
\hline 
BLh & 1.475 & 1.475 & 2.95 & 1.0 & LR & 0.14 & 0.58 & 0.33 & 69.26 & 4.32 & 0.22 & 20.47 & 0.44\\
BLQ & 1.475 & 1.475 & 2.95 & 1.0 & LR & 0.13 & 0.04 & 0.27 & 2.98 & 0.0 & 0.29 & 52.87 & 0.41\\
\hline 
BLh & 1.5 & 1.5 & 3.0 & 1.0 & LR & 0.14 & 0.13 & 0.31 & 13.05 & 0.08 & 0.26 & 28.1 & 0.31\\
BLQ & 1.5 & 1.5 & 3.0 & 1.0 & LR & 0.12 & 0.01 & 0.15 & 0.14 & 0.0 & 0.2 & 78.16 & 0.38\\
\hline 
BLh & 1.6 & 1.6 & 3.2 & 1.0 & LR & 0.01 & 0.01 & 0.18 & 0.21 & 0.0 & 0.21 & 60.81 & 0.29\\
BLQ & 1.6 & 1.6 & 3.2 & 1.0 & LR & 0.0 & 0.0 & 0.17 & 0.14 & 0.0 & 0.22 & 82.65 & 0.34\\
\hline 
BLh & 1.482 & 1.259 & 2.74 & 1.18 & LR & 81.52 & 2.86 & 0.19 & 128.08 & 0.21 & 0.17 & 13.85 & 0.58\\
BLQ & 1.482 & 1.259 & 2.74 & 1.18 & LR & 17.79 & 3.25 & 0.22 & 184.61 & 1.07 & 0.13 & 11.14 & 0.43\\
\hline 
BLh & 1.856 & 1.02 & 2.88 & 1.82 & LR & 60.78 & 7.46 & 0.11 & 109.1 & 0.0 & 0.05 & 4.37 & 0.13\\
BLQ & 1.856 & 1.02 & 2.88 & 1.82 & LR & 60.78 & 7.46 & 0.11 & 102.74 & 0.05 & 0.05 & 4.27 & 0.13\\
\hline 
\hline 
\end{tabular} 
}
\end{center} 
\end{table*}

We summarize the results of this analysis in table \ref{tab:ejecta}. We report the mean ejecta properties from our simulations. When comparing the BLh and the BLQ binaries, we do not find systematic differences in the total ejecta mass, or in the average entropy, composition, or angular distribution. The only robust trend appears in the velocity distribution of the ejecta. For this purpose, we define the fast moving ejecta as the baryonic matter which follows the condition $W \beta > 1$ where $W$ is the Lorentz factor and $\beta = v/c$. The bulk ejecta from both the BLh and the BLQ binaries are subrelativistic, with asymptotic velocities in the range $0.1{-}0.3\ c$, in agreement with previous findings \cite{Bauswein:2013yna, Sekiguchi:2015dma, Radice:2016dwd, Radice:2018pdn, Bernuzzi:2020txg, Vsevolod:2020pak}. However, we also observe that a small fraction of the ejecta (up to $10^{-4} \msun$) achieves asymptotic velocities as large as 0.8~$c$ \cite{Metzger:2014yda, Hotokezaka:2018gmo, Radice:2018pdn, Radice:2018ghv, Nedora:2021eoj}. It is in this latter component of the ejecta that we find a systematic difference between the BLh and the BLQ binaries. In particular, the BLQ binaries that do not undergo prompt collapse produce larger amounts of ejecta with velocity $W \beta > 1$ than the corresponding BLh binaries. The total kinetic energy of this component of the ejecta is also larger for the BLQ EOS. This fast-moving tail of the ejecta is launched when the remnant bounces back after the merger \cite{Radice:2018pdn, Nedora:2021eoj}, so we speculate that the differences between the BLh and BLQ binaries in this component of the outflow is due to the stronger oscillations experienced by the BLQ remnants after merger. We remark that a similar effect was reported by \cite{Radice:2018pdn}. In that case it was the appearance of hyperons to cause the merger remnant to bounce more strongly, while here the stronger bounce of the BLQ binaries is caused by the QCD phase transition.

\begin{figure*}
  \centering
  \includegraphics[width=0.32\textwidth]{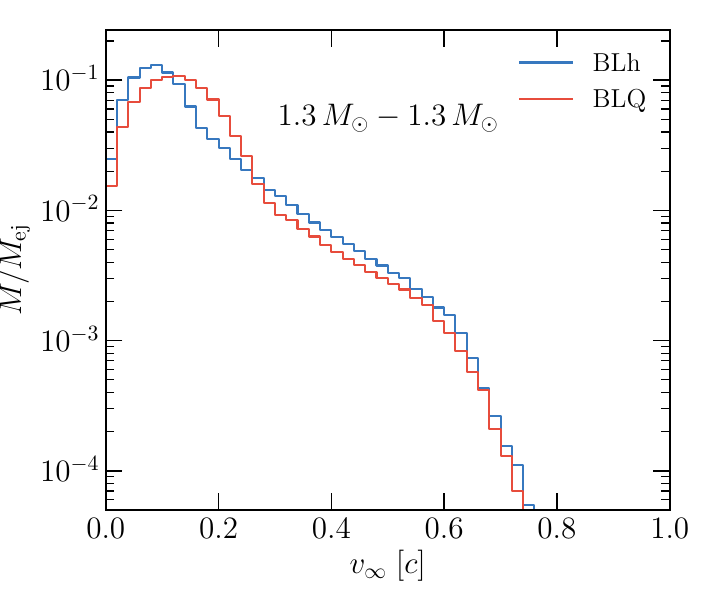}
  \includegraphics[width=0.32\textwidth]{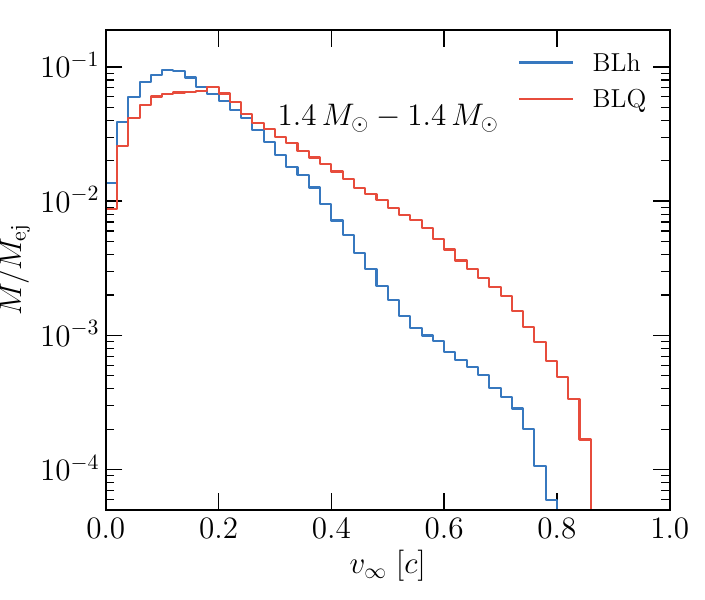}
  \includegraphics[width=0.32\textwidth]{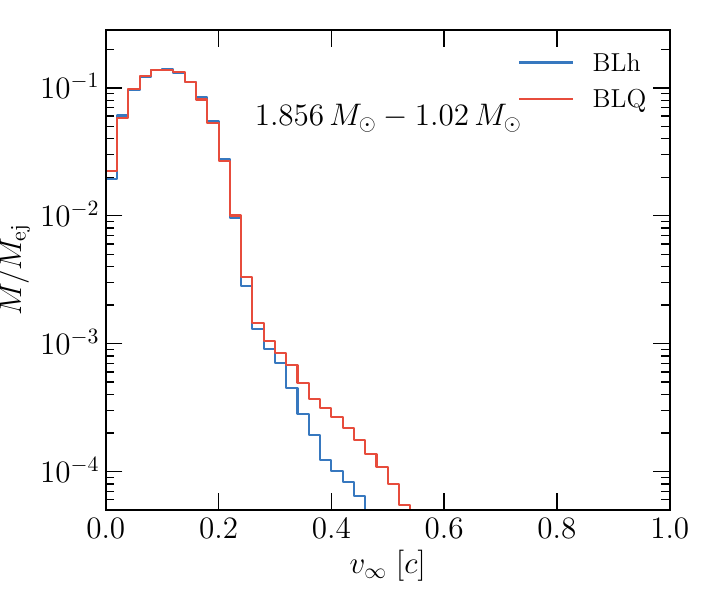}
  \includegraphics[width=0.32\textwidth]{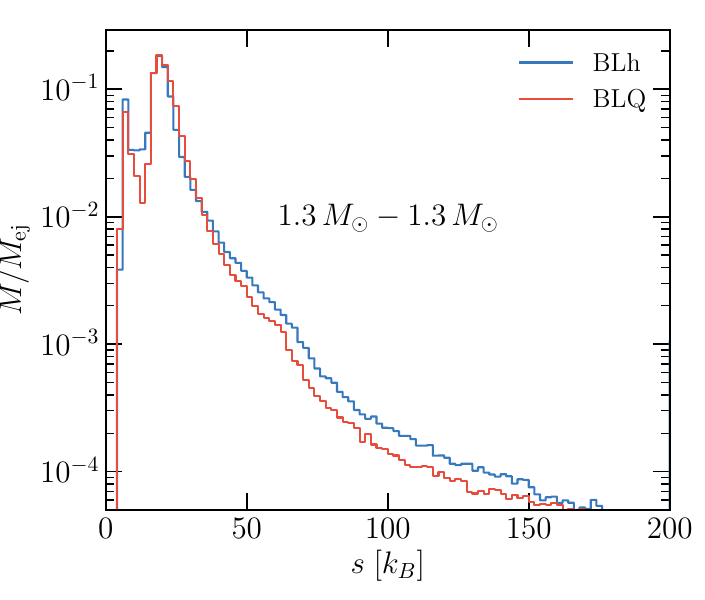}
  \includegraphics[width=0.32\textwidth]{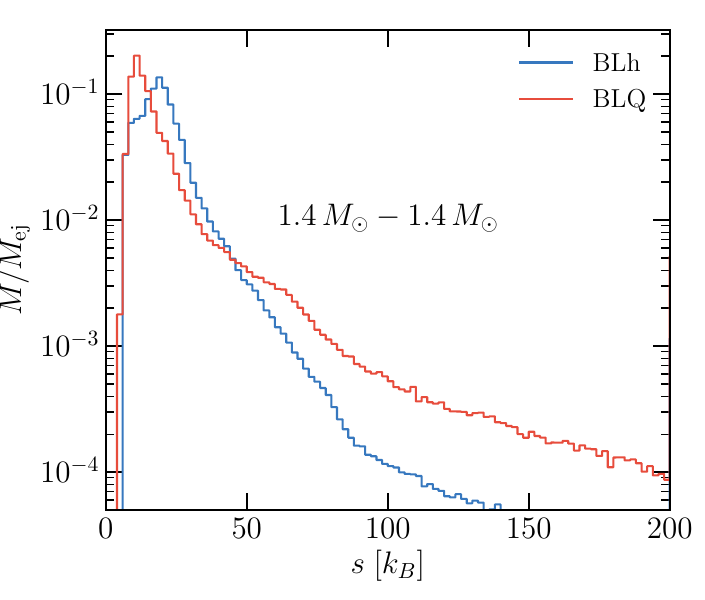}
  \includegraphics[width=0.32\textwidth]{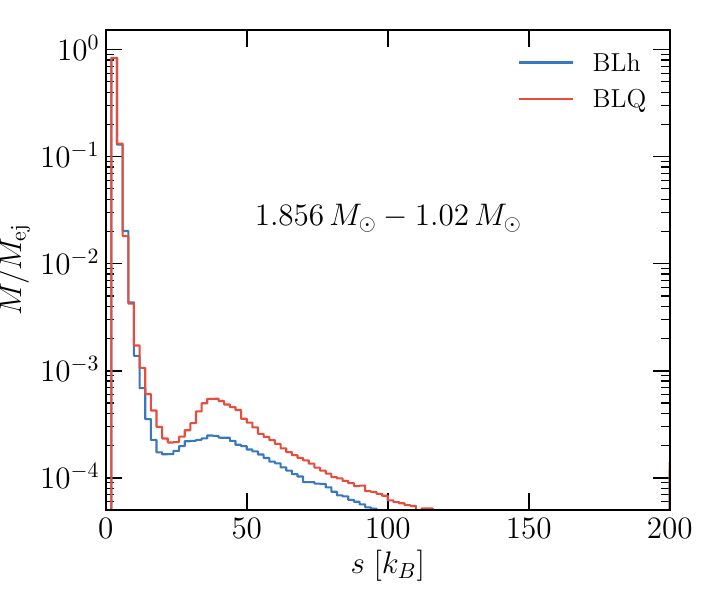}
  \includegraphics[width=0.32\textwidth]{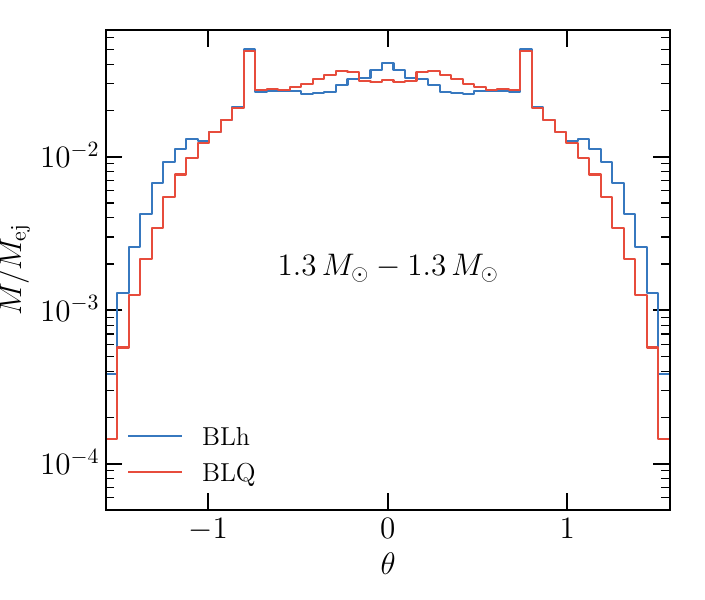}
  \includegraphics[width=0.32\textwidth]{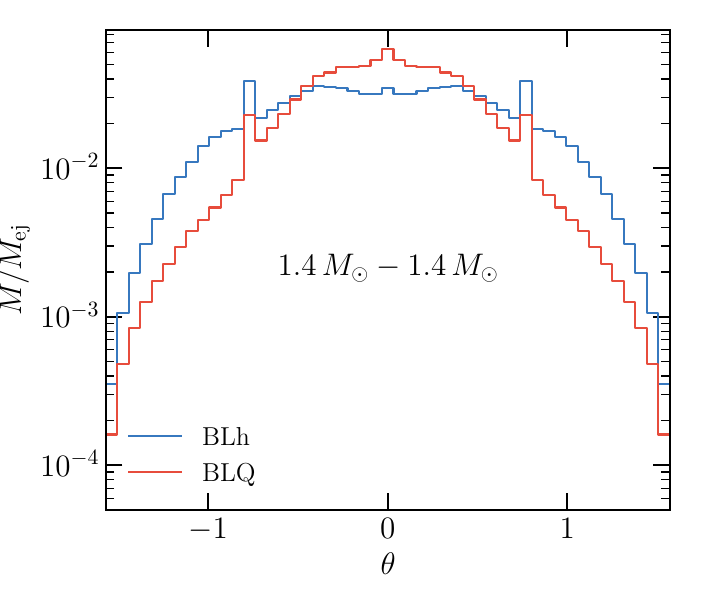}
  \includegraphics[width=0.32\textwidth]{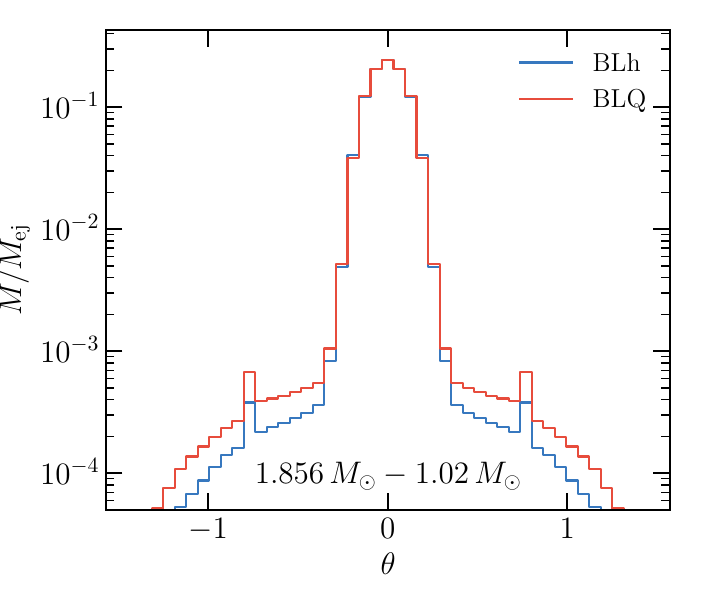}
  \includegraphics[width=0.32\textwidth]{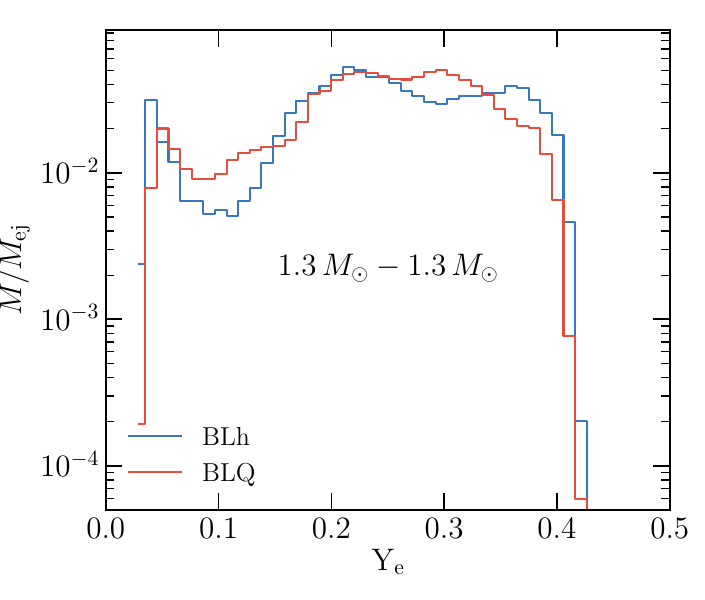}
  \includegraphics[width=0.32\textwidth]{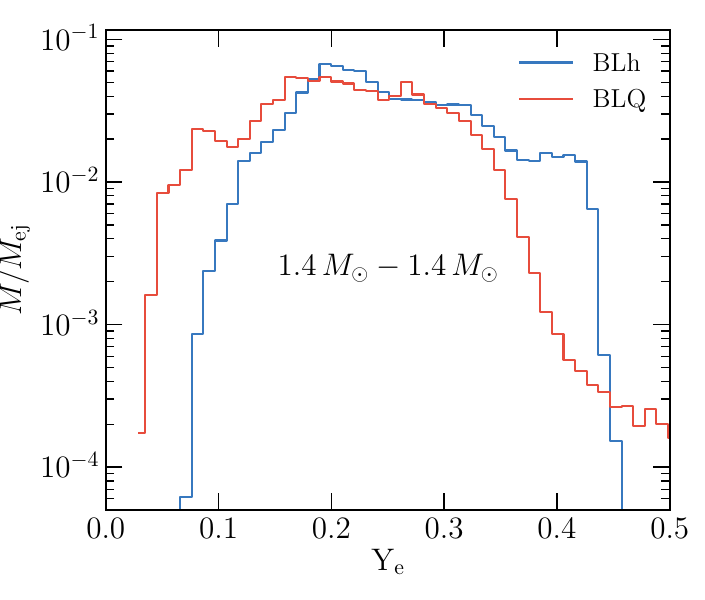}
  \includegraphics[width=0.32\textwidth]{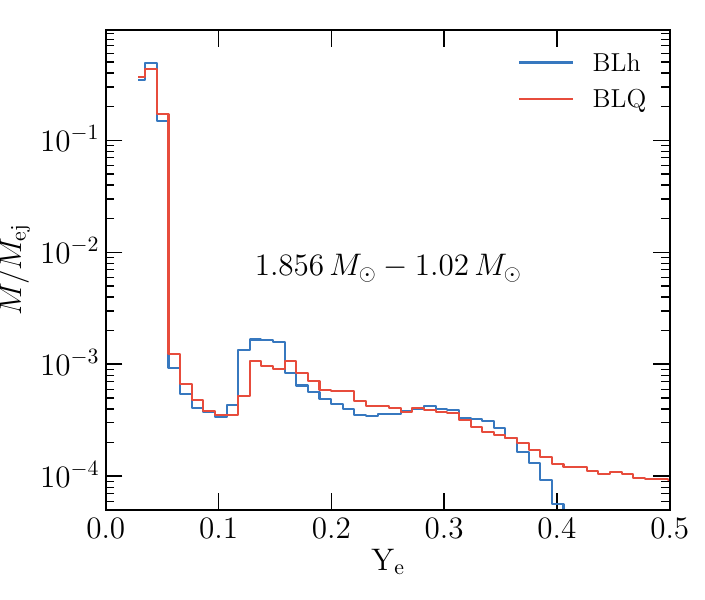}
  \caption{Histograms of the asymptotic velocity $v_{\infty}$, specific entropy $s$, angle with the orbital plane $\theta$, and electron fraction $Y_e$ of the ejecta for three representative binary configurations evolved with the BLh and BLQ EOSs. The most significant differences are seen in the $1.4\ \msun$ - $1.4\ \msun$ binary, for which the BLQ EOS predicts rapid BH formation, while the BLh EOS predicts a long-lived remnant. We note that $M$ here represents the mass of the ejecta in the corresponding bins normalized to $M_{ej}$ i.e. the total ejecta mass as reported in table \ref{tab:ejecta}. }
  \label{fig:ejecta}
\end{figure*}

Figure \ref{fig:ejecta} shows histograms of the properties of the outflows for three representative binaries. The $1.3\ \msun - 1.3\ \msun$ binary is representative of a low-mass merger for which the appearance of deconfined quarks does not lead to qualitative changes in the dynamics during the first milliseconds from the merger. This is due to the relatively low densities achieved by this binary ($\rho_{\max} \lesssim 4\;\rho_{\rm nuc}$) in the first ${\sim}5\ {\rm ms}$ of the merger. As the stars merge, the quark fraction $\rm{Y_q}$, in phase with central density, attains a maximum of 0.3. Further oscillations in density are constrained between $2.8 \rho_{\rm nuc} - 4 \rho_{\rm nuc}$ and are not conducive for the formation of a large amount of deconfined quarks whose fraction oscillates between $0-0.15$ for $t - t_{\rm merg} < 5\ {\rm ms}$. It is important to emphasize that the presence of a phase transition still leads to a qualitative change in the outcome of the $1.3\ \msun - 1.3\ \msun$ merger. Indeed, the BLQ binary collapses to BH after ${\sim}15\ {\rm ms}$ of the merger, while the BLh binary forms a long-lived remnant (see table \ref{tab:GW_property}). However, these differences manifest themselves on somewhat longer timescales than those relevant for the production of the dynamical ejecta, which is launched within ${\sim} 1{-}5\ {\rm ms}$ of the merger \cite{Radice:2018pdn}.

The $1.4\ \msun - 1.4\ \msun$ binary is representative of a binary for which the impact of quark deconfinement leads to dramatic qualitative differences in the dynamics of the merger. The BLh binary forms a long-lived remnant that does not collapse within our simulation time, while the BLQ binary experiences a catastrophic loss of pressure support as hadrons are converted to quarks and forms a BH within ${\sim}2\ {\rm ms}$ of the merger. Neither is a case of prompt BH formation: the $1.4\ \msun - 1.4\ \msun$ BLQ binary still experiences one violent bounce before collapsing. The more violent merger of the BLQ binary is reflected in a significantly larger amount of fast moving ejecta (figure~\ref{fig:ejecta} and Tab.~\ref{tab:ejecta}). This interpretation is confirmed by the presence of a significant excess of high-entropy, shock-heated, material in the BLQ ejecta. The ejecta distribution is also more concentrated close to the orbital plane, as expected for the fast-tail of the shock driven ejecta \cite{Radice:2018pdn, Nedora:2021eoj}. Interestingly, the electron fraction in the ejecta of the $1.4\ \msun - 1.4\ \msun$ BLQ binary is smaller than that of the corresponding BLh binary. This is because the BLh ejecta are irradiated by neutrinos from the massive NS remnant, which is absent in the BLQ binary (due to the early BH formation).

The $1.856\ \msun - 1.020\ \msun$ binary is an example of a merger resulting in prompt BH formation with both the BLh and BLQ EOSs. The dynamical ejecta is entirely driven by tidal torques on the secondary NS, so we do not expect any effect due to the phase transition. Indeed, the differences between the $1.856\ \msun$ - $1.020\ \msun$ BLh and BLQ binaries shown in figure~\ref{fig:ejecta} are not robust with resolution. However, our simulation reveal another interesting effect. The ejecta has two components. A low electron fraction, low entropy component with most of the ejecta mass and a high electron fraction $\mathrm{Y_e} > 0.3$ and high entropy $s \gtrsim 25\ {k_B}$ component. The presence of a second component in the ejecta in highly asymmetric binaries was already reported in Refs.~\cite{Lehner:2016lxy, Sekiguchi:2016bjd, Bernuzzi:2020txg}, where it has been attributed to the presence of a residual shock driven component of the outflows. However, a careful analysis of the evolution of the ejecta in the orbital plane as a function of time suggests that, at least for the binaries considered here, this second component is due to the presence of internal shocks in the tidal debris.

\begin{figure*}[t]
  \centering
  \includegraphics[width=0.49\textwidth]{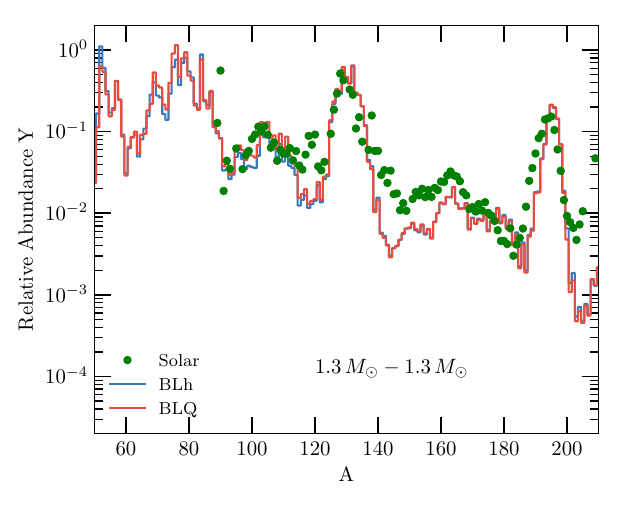}
  \includegraphics[width=0.49\textwidth]{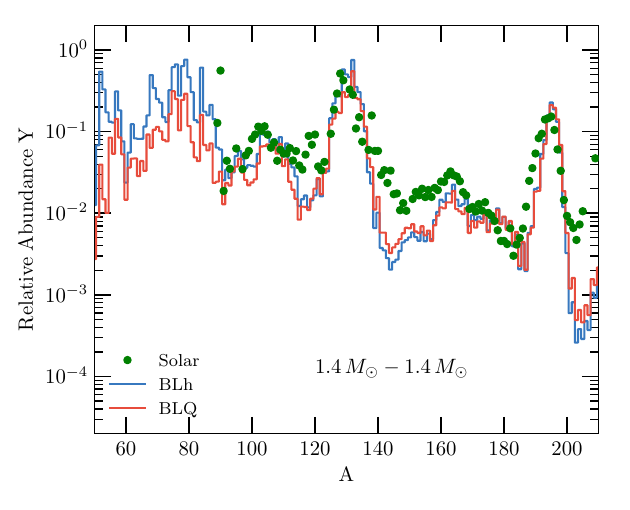}
  \\
  \centering
  \includegraphics[width=0.49\textwidth]{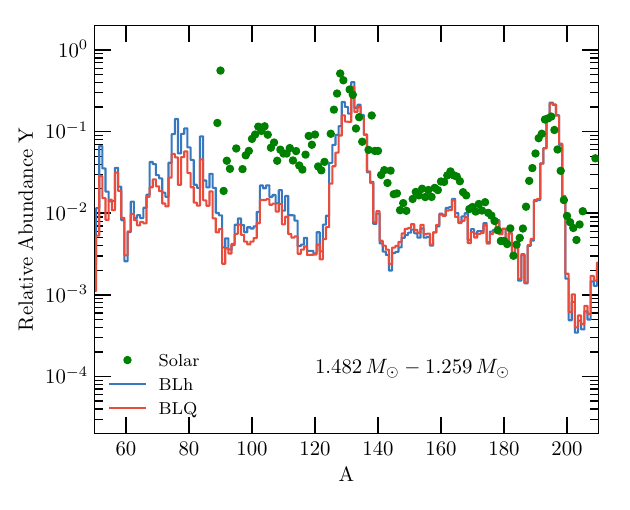}
  \includegraphics[width=0.49\textwidth]{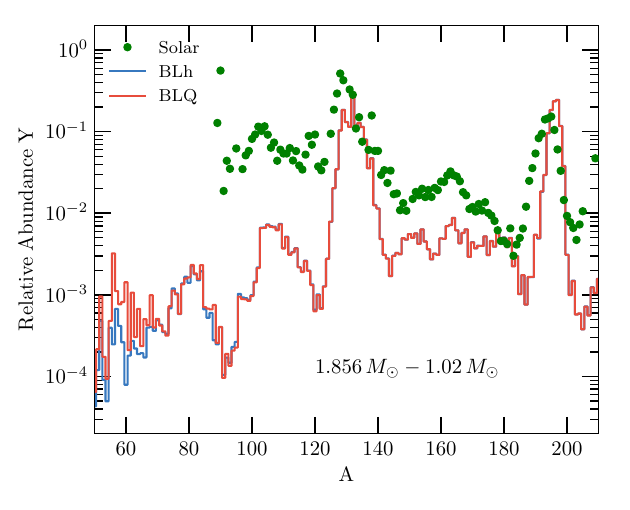}
  \caption{Nucleosynthesis yields of the dynamical ejecta from selected binaries. The final relative abundances in the ejecta are insensitive to the appearance of quarks, but are instead sensitive to the binary mass-ratio. Comparable-mass binaries produce r-process elements with relative abundances close to Solar $r$-process residual, while high-mass ratio binaries show ratios of heavy to light r-process abundances that are significantly larger than the Solar $r$-process residual. We normalize the yields at a given $\rm{A}$ with respect to the yields in the 3$^{rd}$ r-process peak i.e. $\rm{A} \in [180, 200]$ to report the relative abundance Y.}
  \label{fig:nucsyn}
\end{figure*}

The outflow from BNS mergers realizes the conditions for the production of heavy elements via the rapid neutron capture process \cite{Thielemann:2017acv}. We use the methodology described in detail in Refs.~\cite{Radice:2016dwd} and \cite{Radice:2018pdn} to compute the relative abundances of heavy nuclei produced in the dynamical ejecta from our simulations. Our results are shown in figure~\ref{fig:nucsyn}. We find that the presence of deconfined quarks in the BLQ binaries does not leave a significant imprint on their nucleosynthesis yields. Even in the case of the $1.4\ \msun - 1.4\ \msun$ binaries, for which the phase transition has a strong impact on the merger dynamics, we find that the differences in the yields are only modest. The variation in the relative elemental abundances in the ejecta as the mass ratio of the binary is varied, is significantly larger. Indeed, we find that while the dynamical ejecta from binaries with mass ratio $q \simeq 1$ robustly produce elements with relative abundances close to Solar $r$-process residual, the higher mass ratio mergers tend to overproduce second and third r-process peak elements. This is because asymmetric binaries produce a larger amount of neutron rich, cold, tidal ejecta \cite{Radice:2018pdn, Nedora:2021eoj}.

\subsection{Remnant Disks}
\label{subsec:Disk}

\begin{figure*}[!ht]
\includegraphics[width=\textwidth]{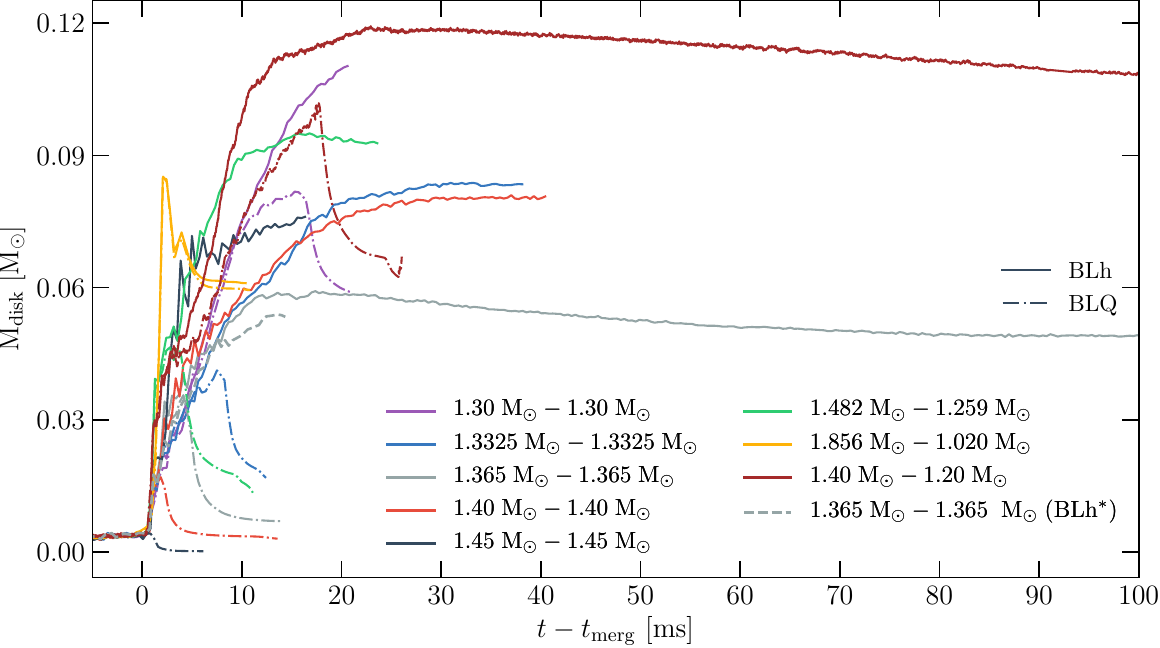}
\caption{\label{fig:disk_mass} Evolution of disk mass for a subset of our simulations. Binaries with the BLh EOS (solid lines) form stable, long-lived remnants with disks evolving on long timescales. The binaries with deconfined quarks (dotted lines) result in the formation of BHs. The gravitational collapse is accompanied by the accretion of a significant fraction of the disk over a timescale of few milliseconds. Binaries where remnants from both EOSs undergo prompt collapse do not show significant differences in their disk mass evolution.}
\end{figure*}

Following the conventions in \cite{Radice:2017lry}, we define the accretion disk as all baryonic matter with rest mass density $\rho < 10^{13} \rm{g\;cm^{-3}}$. We remark that this density threshold corresponds to the approximate location where the angular velocity of the remnant becomes Keplerian \cite{Hanauske:2016gia}. In the case of BH remnants, all of the matter outside of the BH has $\rho < 10^{13} \rm{g\; cm^{-3}}$. Furthermore, we exclude all matter enclosed by the BH apparent horizon i.e. we only include regions where $\alpha \geq 0.3$. In other words,
\begin{equation}
    M_{\rm disk} =\int \sqrt{\gamma}\;W\;\rho\;dx\;dy\;dz
\end{equation}
where $\gamma$ represents the determinant of the spatial metric and $W$ is the Lorentz factor. The integration is carried out over all matter in the region defined above.

As discussed in detail in Radice et al.~\cite{Radice:2018pdn}, the remnant accretion disk is formed of material that is squeezed out of the collisional interface between the NSs during and shortly after the merger. So the disk mass initially increases with time, as shown in figure~\ref{fig:disk_mass}. For those binaries that do not form BHs, the disk mass peaks within $10{-}20\ {\rm ms}$ of the merger. Over longer timescales the angular momentum transport due to spiral density waves drives accretion and a secular outflow from the disk \cite{Nedora:2019jhl, Vsevolod:2020pak}. Angular momentum transport due to MHD turbulence is also expected to contribute to this process, however our simulations did not include magnetic fields, so they cannot describe this phenomenon. At the same time, we remark that in our previous study we found that in the first ${\sim}100\ {\rm ms}$ of the merger the spiral waves are expected to be the dominant mechanism for angular momentum transport \cite{Nedora:2019jhl}.

The dynamics is very different for binaries that form BHs. When the central part of the remnant collapses a significant fraction of the disk is accreted within a few milliseconds (figure~\ref{fig:disk_mass}), as also reported in Ref.~\cite{Vsevolod:2020pak}. Since the BLQ EOS predicts early BH formation for all binaries considered in this study, while most of the BLh binaries form long-lived remnant, this process leads to significant differences between the remnant disks for the BLh and BLQ binaries. Exceptions to these are the massive equal-mass binaries that collapse promptly for both the BLh and BLQ EOS resulting in a rapid disk accretion post merger and the $1.856\ \msun - 1.020\ \msun$ system, for which the disk is formed from the tidal disruption of the secondary NS in the late inspiral, prior to the production of a significant amount of deconfined quarks.

\section{EM Signatures}
\label{sec:em_signatures}

\subsection{Kilonova Light Curves}
\label{subsec:kilonova_light_curves}

\begin{figure}
    \includegraphics[width=0.49\textwidth]{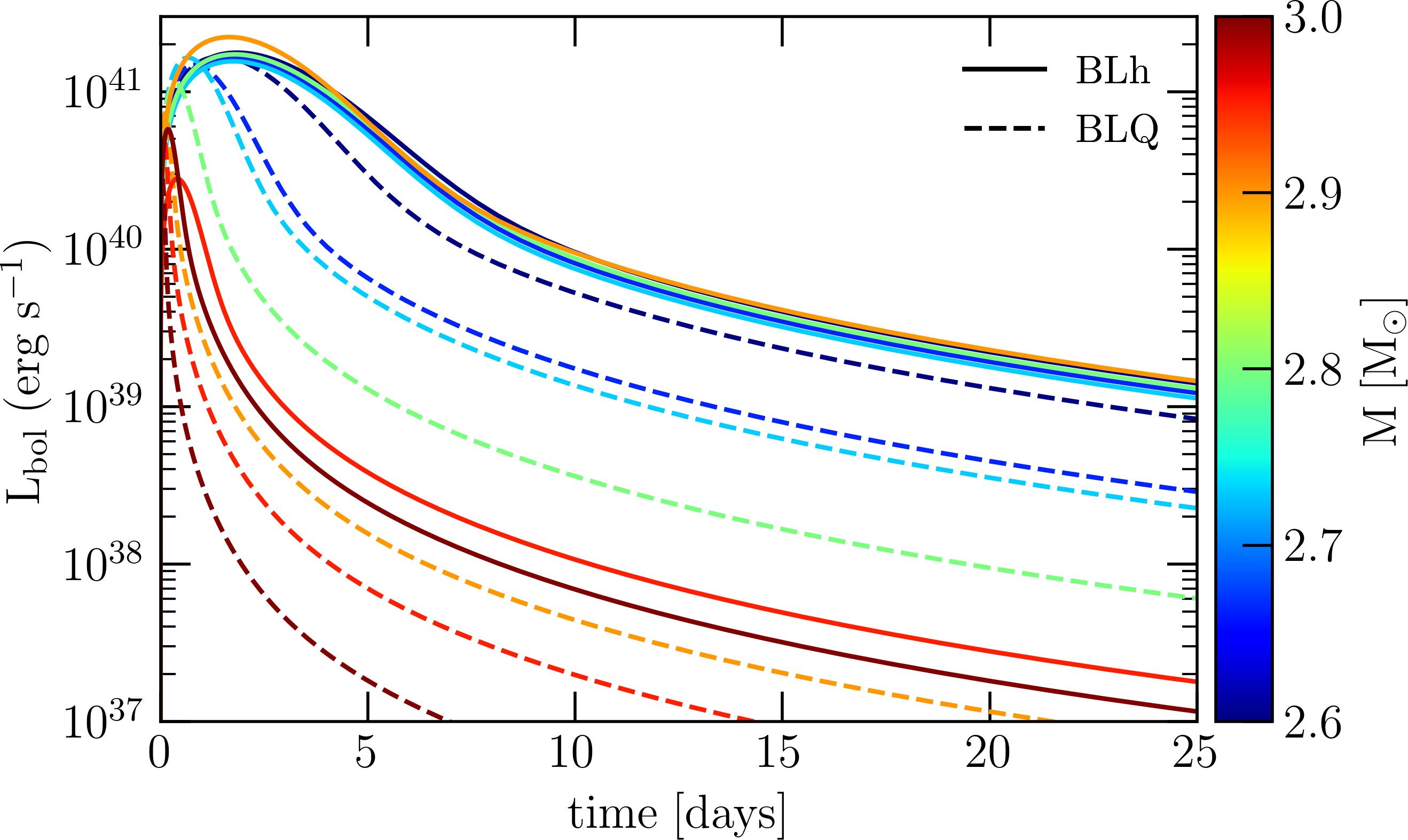}
    \caption{
        Kilonova light curves for a subset of our simulations with $q = 1$. The colour code represents the total mass of the binary with the dashed (solid) curves indicating models with (without) a QCD phase transition. In general, BLh binaries are more luminous and the brightness decreases with increasing mass.
    }
    \label{fig:knova_lcs}
\end{figure}
For the purpose of computing kilonova light curves, we compute the ejecta properties using the Bernoulli criterion, which allows us to include both the dynamical ejecta discussed above and the wind ejecta emerging at later times in our simulations \cite{Kastaun:2014fna, Bovard:2017mvn, Nedora:2019jhl}. We also assume that $20\%$ of the disk mass at the end of the simulation is unbound by winds \cite{Fujibayashi:2017puw, Fernandez:2018kax, Fernandez:2020oow}. For the wind component of the ejecta, we assume mass-averaged electron fraction and velocity to be $0.2$ and $0.1c$ respectively \cite{Fernandez:2013tya}. We calculate bolometric luminosities using a single component kilonova light curve model, whose basic equations are given in Refs.~\cite{Arnett:1982ioj, Chatzopoulos:2011vj, Villar:2017wcc}. In addition to the total ejecta mass, we also extract from the simulations, the mass-averaged velocity and electron fraction of the outflow. The latter is used to calculate the matter opacities using an analytic fit to the results of Ref.~\cite{Tanaka:2019iqp}. The input heating rate and thermal efficiency of the ejecta used here are taken from Refs.~\cite{Korobkin:2012uy} and \cite{Barnes:2016umi}, respectively.

We show bolometric lightcurves for models with $\rm{q = 1}$ in figure \ref{fig:knova_lcs}. Generically, we find that the BLh binaries lead to more luminous and slowly evolving kilonovae than their counterparts with the BLQ EOS. However, these differences are entirely attributable to the differences in life time between the BLh and BLQ remnants. Indeed, we also observe that the kilonova becomes dimmer for more massive binaries. Equal mass binaries undergoing prompt BH formation have very small ejecta and disk masses, owing to which their kilonovae are very dim. Kilonova observations are powerful probes to determine the life time of BNS merger remnants \cite{Radice:2020ddv}. However, the time to collapse for a merger remnant might depend on factors other than a phase transition such as particular features in the hadronic EOS or magnetic and neutrino effects \cite{Radice:2018xqa}.

We observe that the kilonova corresponding to the binary merger of $1.45\ \msun - 1.45\ \msun$ with the BLh EOS is the brightest among all our models as shown in figure \ref{fig:knova_lcs}. This may be attributed to the fact that the ejecta from this model is the most energetic (see table \ref{tab:ejecta}) of all our systems. It is important to note that the binary is very close to the prompt collapse threshold for the BLh EOS and hence is subjected to large uncertainties arising from spatial resolution which also levies large uncertainties in the calculation of disk masses. Indeed, we observe that at standard resolution (SR) the binary is long-lived with a lifetime of $\gtrsim 14 \;\rm{ms}$ (table \ref{tab:GW_property}). However, with the low resolution (LR) grid setup, it collapses to a BH on a much shorter time scale of $\sim 3\; \rm{ms}$. Another aspect to note here is that this is the only binary where the BLQ remnant collapses promptly to a BH but the BLh remnant does not. 

\subsection{Kilonova Afterglow}
\label{subsec:kilonova_afterglow}

The fast-moving tail of the ejecta is expected to drive shocks in the ISM which might produce synchrotron radiation over a wide range of the EM spectrum over a timescale of months to years from the merger. This is the so-called kilonova afterglow \cite{Nakar:2011cw, Hotokezaka:2015eja, Hotokezaka:2018gmo, Margalit:2020bdk, Nedora:2021eoj}. This scenario has been invoked to explain the recent deviation of the X-ray afterglow in GW170817 from the theoretical expectations for a relativistic jet
\cite{Hajela:2021faz, Nedora:2021eoj}, although other interpretations of this data are not excluded \cite{Hajela:2021faz, Ishizaki:2021vsx}. The properties of the \kna{} depend sensibly on the ejecta mass and velocity distributions. It is therefore conceivable that late time observations of BNS mergers could probe the presence of phase transitions in the EOS of dense matter.

We calculate the \lcs{} of the synchrotron radiation arising from the interaction between the dynamical ejecta and the ISM with the semi-analytic code \texttt{PyBlastAfterglow} \citep{Nedora:2021eoj,Hajela:2021faz}. The code computes the synchrotron radiation that arises from electrons accelerated in the amplified magnetic field in the forward shock, \ie, in the shock between the expanding blast wave and ISM. The total flux density is computed by integrating the flux from each element of the solid angle over equal-time arrival surfaces. The ISM is assumed to be cold and uniform with density $n_{\rm ISM}$. The equipartition microphisical parameters, describing the energy conversion efficiency between the shock and the magnetic fields and electrons, $\epsilon_e$ and $\epsilon_B$ respectively, are assumed to be constant. The initial conditions for the code are given by the kinetic energy and angular distribution of the ejecta from the merger simulations. Its evolution is computed assuming only adiabatic energy losses and no lateral spreading.

We set the free parameters as follows. The observational angle, namely the angle between the line of sight and the polar axis of the BNS system, is $\theta_{\text{obs}}=30\,$deg, which is consistent with the observational geometry for GW170817 \cite{TheLIGOScientific:2017qsa}. We consider a source at $40\,$Mpc with the redshift $z=0.0099$. The ISM density and microphysical parameters are set as $n_{\rm ISM}\in(10^{-3}, 10^{-2})\,$cm$^{-3}$, $p = 2.15$, $\epsilon_e=0.2$, and $\epsilon_B=5\times10^{-3}$. These values are chosen from the respective credibility intervals inferred for GRB170817A \cite{Hajela:2019mjy}. Note however, that the kilonova afterglow might have different microphysical parameters as compared to the gamma ray burst (GRB) afterglow. Indeed, recent observations suggest the onset of the spectral evolution of the synchrotron emission from GW170817 \cite{Hajela:2021faz}.

\begin{figure}
    \includegraphics[width=0.49\textwidth]{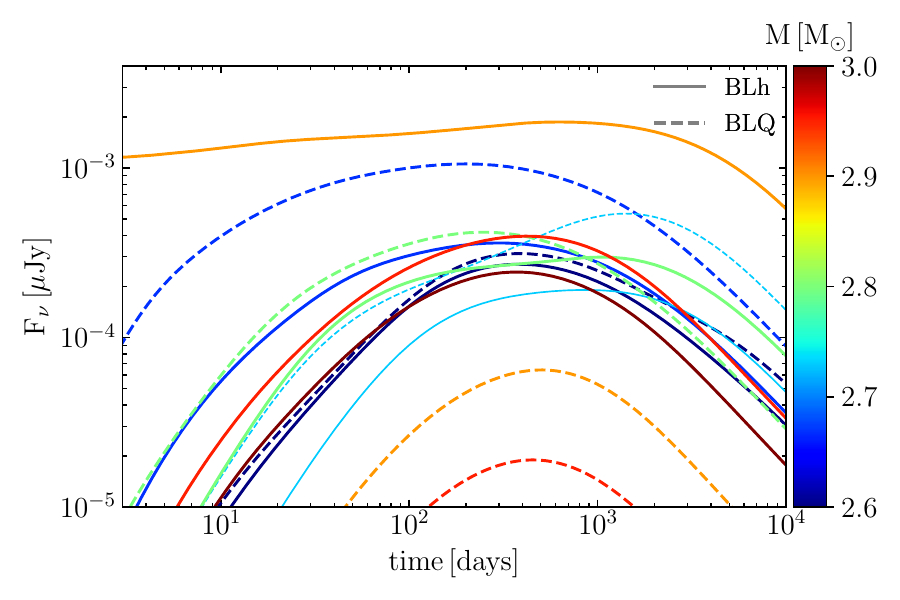}
    \caption{
        Kilonova afterglow \lcs{} at $1$~keV for a set of equal mass models. The models' total mass is color-coded. Dashed (solid) curves indicate models with (without) phase transition. The plot shows that the afterglow of models with phase transition in general is brighter and more extended in time.
    }
    \label{fig:afg_lcs}
\end{figure}

\begin{figure*}
    \includegraphics[width=0.49\textwidth]{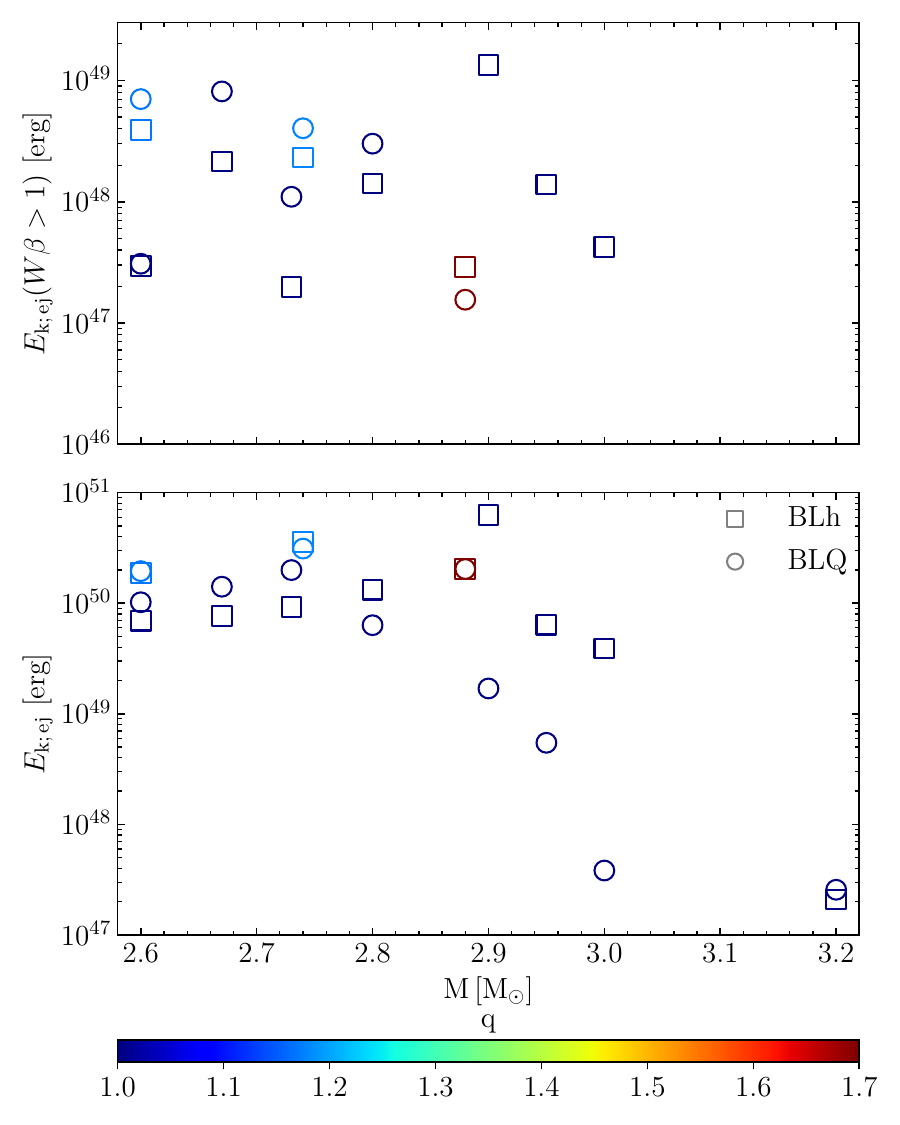}
    \includegraphics[width=0.49\textwidth]{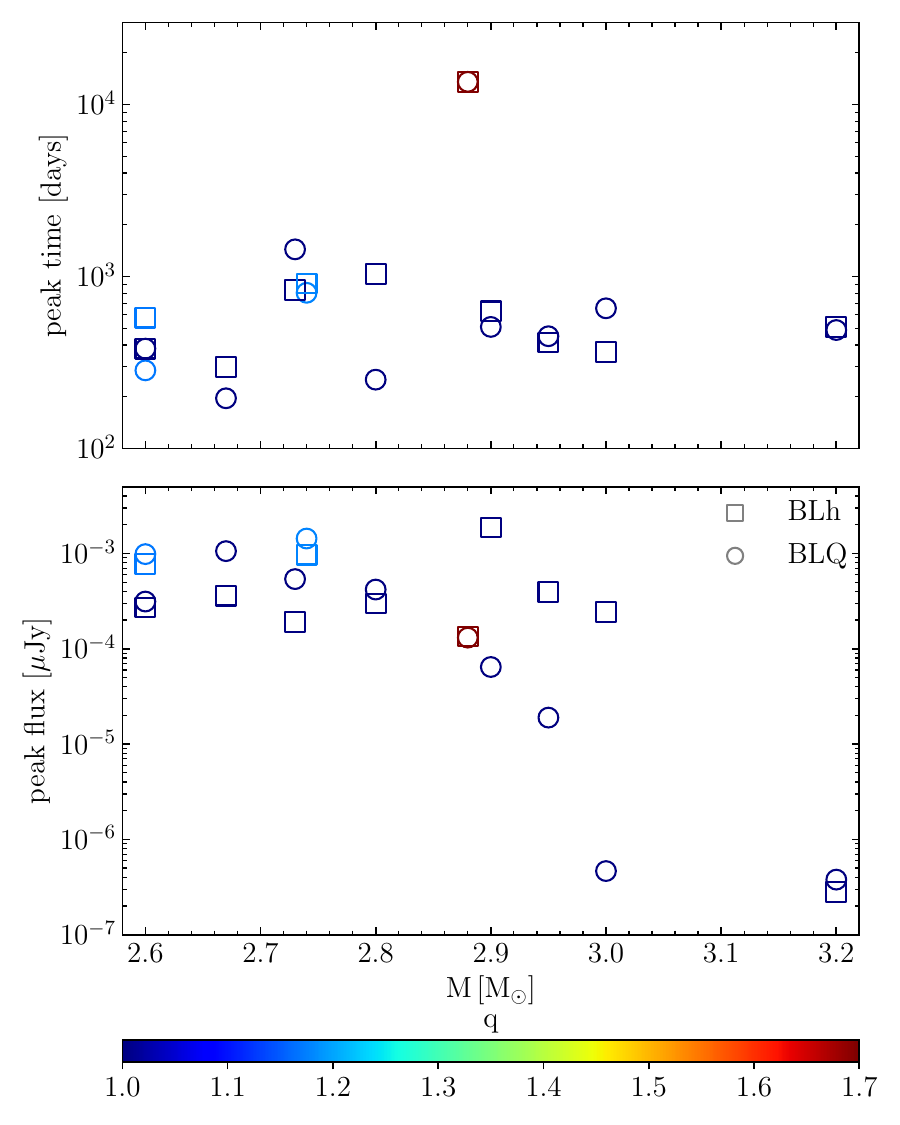}
    \caption{
        The ejecta kinetic energy (\emph{left two panels}) and \kna{} properties (\emph{right two panels}) for the simulations with and without phase transition (BLQ and BLh, respectively). The kinetic energy is shown separately for the entire ejecta (\emph{lower subpanel}) and for the fast component only (\emph{upper panel}). The \kna{} properties are the \lcs{}' peak time (\emph{upper panel}) and flux (\emph{lower panel}). Circle (squre) markers indicate models with (without) phse transition, \ie, models with BLQ (BLh) EOS. The plot shows a correlation between the peak flux and the total kinetic energy.        The effect of the phase transition is very prominent at high mass binaries, where        the softening of EOS leads to prompt collapse, reducing the ejecta kinetic energy and
        peak flux.
    }
     \label{fig:afg_and_ek_scatter}
\end{figure*}

Figure \ref{fig:afg_lcs} shows the \kna{} \lcs{} from a set of representative equal mass models with total mass ranging from $2.6\,\Msun$ to $3.0\,\Msun$. Notably, the \kna{} properties of low mass models with BLh EOS are rather independent of $\rm{M}$, peaking at ${\lesssim}10^3\,$days and reaching flux density ${\simeq}10\,\muJy$. The inclusion of a phase transition in these models, generally, leads to broader and slightly brighter \kna{} \lcs{}, as the fast tail of the ejecta of these models becomes more energetic (as discussed previously and as shown in the left panel of figure~\ref{fig:afg_and_ek_scatter}). This is especially apparent for the model with $\rm{M}=2.67\,\Msun$, where the inclusion of the phase transition leads to a considerably more energetic fast ejecta tail (see table~\ref{tab:ejecta} and upper left panel of figure~\ref{fig:afg_and_ek_scatter}), which in turn leads to a significantly broader \lc{}.

The general properties of the \kna{} \lcs{}, \ie, the peak flux $\Fp$ and the peak time time $\tp$ are shown in figure~\ref{fig:afg_and_ek_scatter} (right panel) for all models. Notably, $\Fp$ is well correlated with $\Ek$, a trend that is much less clear in simulations with large mass ratio \cite{Nedora:2021eoj}. The plot shows that indeed, among the low mass models, the inclusion of the phase transition raises $\Fp$. However, a difference of  ${\leq}10\,\muJy$ is smaller than the systematic uncertainties introduced by ill-constrained microphysical parameters. When more massive models, $\rm{M}\in(2.9,3.1)\,\Msun$, are considered, the difference in $\Fp$ becomes larger than the systematic uncertainties, as models with BLQ EOS undergo prompt collapse, producing less energetic ejecta.

The effect of the phase transition on the peak time is, however, unclear. The largest difference is observed for the model with $1.4\ \msun - 1.4\ \msun$, where the \lc{} $\tp$ of the model without phase transition is ${\sim9}$ times larger. For the other models the difference is within the systematic uncertainties due to ill-constrained $n_{\rm ISM}$. As was the case for early-time kilonova, here too we observe the $1.45\ \msun - 1.45\ \msun$ binary with the BLh EOS to have the brightest overall afterglow. For this binary, the large kinetic energy of the ejecta is responsible for the early rise and subsequently large values of the synchrotron flux. 

\section{Conclusions}
\label{sec:conclusions}

In this work we presented fully general relativistic neutrino-radiation hydrodynamics simulations of BNS mergers with a first order phase transition to deconfined quark matter. We considered and systematically analysed a wide range of BNS configurations. We studied their evolution using two EOSs with identical hadronic physics but that, respectively, included or excluded a QCD phase transition to deconfined quarks: the BLQ EOS and BLh EOS. The BLQ EOS is presented here for the first time. By comparing the results obtained with both the EOSs, we computed the observable signatures: GW, EM counterparts, and nucleosynthesis yields resulting from the phase transition.

The BLh EOS, which describes the hadronic phase of dense stellar matter, has been calculated making use of a finite temperature BHF approach \cite{1993PhLB..311....9B,1994PhR...242..165B,1999PhRvC..59..682B} starting from modern two-body and three-body nuclear interactions derived within ChEFT. To model the quark phase we used a phenomenological bag model EOS which includes the effects of gluon mediated QCD interactions between quarks up to the second order in the QCD coupling $\alpha_s$. We assumed a first order transitions between the two phases and using the Gibbs construction \cite{Glendenning:1992vb} we derived the EOS for hadronic-quark mixed phase. When considering non-spinning isolated NSs, the differences in properties of matter modelled by the zero temperature version of the two EOSs, BLh and BLQ, begin to manifest for stars heavier than $\sim$ $1.7\msun$.

We find that the hot interface region created when the two NSs of a binary system start to merge is the first site where deconfined quark matter can be produced. In this region matter crosses the phase boundary at intermediate densities ($\rho \sim 2\ \rho_{\rm nuc}$) and temperatures ($\rm {T} \sim 30\ \rm {MeV}$). As the merger proceeds and the cores of the two stars fuse, the inner core of the remnant is nearly-adiabatically compressed to high densities ($\rho \gtrsim 3\ \rho_{\rm nuc}$) and crosses the phase boundary at low temperatures (${\rm T} \sim 5\ {\rm MeV}$). The phase transition results in a loss of pressure support in the merger remnants. In particular, the BLQ remnants become more compact and collapse to BH significantly earlier than the corresponding BLh remnants, which do not model the QCD phase transition. These results are in good qualitative agreement with the findings of Most et al.~\cite{Most:2018eaw, Most:2019onn}. Additionally, we find that the threshold mass above which prompt BH formation takes place is lowered by the inclusion of the phase transition in the BLQ EOS.

We employ Lagrangian tracer particles to record the thermodynamic evolution of the binaries by tracking the properties of individual fluid elements. We find that fluid elements repeatedly cross the phase boundary between the hadronic phase and the mixed quark phase as the remnants oscillate. Such dynamics was anticipated by Hanauske et al.~\cite{Hanauske:2019qgs}, but it is shown here for the first time in the context of self-consistent simulation with a first order phase transition. Our analysis shows that BNS mergers probe a large region of the QCD phase diagram, with matter potentially crossing the phase boundary over a large range of temperatures and densities.

The QCD phase transition is most strongly imprinted in the postmerger GW signal. Owing to the rapid softening of the EOS caused by the phase transition, remnants evolved with the BLQ EOS are more compact than the corresponding BLh remnants. This influences the GW signal in two ways. First, because BLQ remnants have a larger gravitational binding energy in absolute value, they radiate a comparably larger amount of GW energy compared to the BLh binaries up to BH formation, at which point the GW emission terminates. Second, the change in the moment of inertia of the remnants due to the phase transition manifests itself as a shift in the postmerger frequency as observed in the GW spectra. While the presence of these shifts appears to be a robust feature of the phase transition \cite{Bauswein:2018bma, Bauswein:2020ggy, Blacker:2020nlq}, we find that for the EOS models we are considering their potential for GW astronomy is limited. On one hand, the magnitude of these shifts is found to be small. On the other hand, because of the early BH formation present in all binaries with phase transitions, the nominal uncertainty with which frequencies can be determined using Fourier analysis can be larger than the magnitude of the frequency shifts themselves. This means that a clean detection of the frequency shift for several of the binaries considered here would be impossible even in the limit of infinite SNRs. This issue is particularly severe for high mass binaries, for which BH formation occurs early. Lower mass binaries show measurable frequency shifts with $\Delta f_2 \lesssim 200\ {\rm Hz}$. However, these deviations are not large compared to those normally present between different hadronic models. Indeed, we find that the postmerger peak frequencies of the binaries with first order phase transition are consistent with the quasi-universal relations that hold for hadronic EOSs \cite{Breschi:2019srl}.

We find the bulk properties of the dynamical ejecta, total mass, average composition and entropy, and geometry, are insensitive to the presence of a phase transition. Indeed, significant differences between the dynamical ejecta of the BLh and BLQ binaries are only present for those binaries that undergo prompt collapse according to the BLQ EOS, but not according to the BLh EOS. However, such differences are also expected when comparing purely hadronic EOSs with different prompt collapse mass thresholds \cite{Radice:2018pdn}, so this is not a genuine signature of phase transition. On the other hand, we find that the BLQ binaries generically produce a larger amount of ejecta with asymptotic velocities exceeding $0.6 c$. This is due to the stronger bounce experienced by such binaries as a result of the phase transition.

We estimated the final abundances of different nuclear species in the dynamical ejecta arising from $r$-process nucleosynthesis. We find no significant difference between the BLh and the BLQ ejecta. This is not surprising given that the properties of the dynamical ejecta between the two set of simulations are very similar. Instead, we find that the nuclear abundances are sensitive to the mass ratio of the binaries. The elemental abundances from comparable mass ratio binaries are instead close to the Solar $r$-process residual.
Higher mass ratio binary generate more tidally driven, neutron rich outflows, which preferentially produce heavy $r$-process elements ($A \gtrsim 130$) \cite{Radice:2018pdn, Bernuzzi:2020txg, Vsevolod:2020pak}.

The remnant accretion disks for the BLQ and BLh binaries have significantly different masses at the end of our simulations. This is due to the fact that all the BLQ remnants form BHs within a short time of the merger (${\lesssim}20\ {\rm ms}$). On the one hand, this terminates the process that leads to the formation of the disk in comparable mass ratio binaries: the shedding of hot material from the newly formed massive NS. On the other hand, BH formation is immediately followed by the rapid accretion of a significant portion of the disk. In contrast, many of the BLh binaries we have considered result in the formation of long-lived remnants. Thus, the variations in the remnant disk mass are entirely explained by the different life times of the BLh and the BLQ BNS merger remnants. Indeed, different hadronic models can also show large variations in the collapse times and the disk masses for BNS merger remnants \cite{Radice:2017lry}.

Similarly, although substantial differences are found between the BLh and BLQ bolometric kilonova light curves, it will be challenging to use UVOIR observations of kilonova events to constrain phase transitions. This is because the variations in kilonova properties observed in our simulations can be produced by effects others than phase transitions. For example, the appearance of hyperons, or the presence of strong magnetic fields \cite{Radice:2018pdn}. More work is needed to understand whether these effects can be disentangled.

The fast-moving tail of the dynamical ejecta is expected to interact with the ISM and produce synchrotron radiation, the so-called kilonova afterglow. We find that, owing to the larger amount of fast ejecta, the BLQ binaries typically produce brighter synchrotron remnants than the BLh binaries. However, if prompt BH formation occurs, then the fast-moving tail of the ejecta is significantly reduced in mass and the synchrotron emission is suppressed. For this reason, the trend is reversed for binaries that undergo prompt BH formation according to the BLQ EOS, but not according to the BLh EOS. Overall, we conclude that kilonova afterglows are a promising avenue to probe a phase transition. Unfortunately, given the large uncertainties in the microphysics of the interaction between the ejecta and the ISM and the accuracy limitations of current simulations, our results cannot be used to quantitatively constrain the presence of phase transition with past or future observations. More work is needed to address these shortcomings.

Our study considered only one hadronic EOS and a specific model for the treatment of the quark phase and of the phase transition. Follow up studies should extend this work to include more hadronic models and different approaches to construct QCD phase transitions. For example, to compare the Gibbs and the Maxwell constructions. This will be the object of our future work.

\begin{acknowledgments}
It is a pleasure to acknowledge Matteo Breschi for having provided updated data for figure~\ref{fig:fit}.
NR simulations were performed on Bridges, Comet, Stampede2 (NSF XSEDE allocation TG-PHY160025), NSF/NCSA Blue Waters (NSF  AWD-1811236) supercomputers. Computations for this research were also performed on the Pennsylvania State University’s Institute for Computational and Data Sciences’ Roar supercomputer. This research used resources of the National Energy Research Scientific Computing Center, a DOE Office of Science User Facility supported by the Office of Science of the U.S.~Department of Energy under Contract No.~DE-AC02-05CH11231.
DR acknowledges support from the U.S. Department of Energy, Office of Science, Division of Nuclear Physics under Award Number(s) DE-SC0021177 and from the National Science Foundation under Grant No. PHY-2011725.
SB acknowledges support by the EU H2020 under ERC Starting Grant, no.~BinGraSp-714626. 
\end{acknowledgments}

\bibliography{bibliography}

\end{document}